\documentclass[acmtog]{acmart}
\setcopyright{rightsretained} 
\acmJournal{TOG}
\acmYear{2022}\acmVolume{41}\acmNumber{4}\acmArticle{42}\acmMonth{7} \acmDOI{10.1145/3528223.3530103}

\usepackage{lipsum}
\usepackage[linesnumbered,ruled,vlined]{algorithm2e}
\usepackage{cleveref}
\newtheorem{theorem}{Theorem}
\usepackage{wrapfig}
\usepackage{enumitem}
\usepackage{multirow}
\usepackage{caption}
\usepackage{subcaption}
\usepackage{tablefootnote}

\usepackage{colortbl}
\definecolor{mygray}{gray}{.9}

\SetKwInput{KwInput}{Input}               
\SetKwInput{KwOutput}{Output}            
\SetKw{Continue}{continue}

\DeclareMathOperator*{\argmax}{arg\,max}

\SetCommentSty{mycommfont}

\crefname{algocf}{alg.}{algs.}
\Crefname{algocf}{Algorithm}{Algorithms}

\AtBeginDocument{
  \providecommand\BibTeX{{
    \normalfont B\kern-0.5em{\scshape i\kern-0.25em b}\kern-0.8em\TeX}}}

\begin{document}
\title{Approximate Convex Decomposition for 3D Meshes with Collision-Aware Concavity and Tree Search}
\author{Xinyue Wei}
\authornote{Both authors contributed equally to this research.}
\email{xiwei@ucsd.edu}
\author{Minghua Liu}
\authornotemark[1]
\email{minghua@ucsd.edu}
\affiliation{
  \institution{University of California San Diego}
  \country{USA}
}

\author{Zhan Ling}
\affiliation{
  \institution{University of California San Diego}
  \country{USA}
}
\email{z6ling@eng.ucsd.edu}

\author{Hao Su}
\affiliation{
  \institution{University of California San Diego}
  \country{USA}
}
\email{haosu@eng.ucsd.edu}

\renewcommand{\shortauthors}{Wei and Liu, et al.}

\begin{abstract}
Approximate convex decomposition aims to decompose a 3D shape into a set of almost convex components, whose convex hulls can then be used to represent the input shape. It thus enables efficient geometry processing algorithms specifically designed for convex shapes and has been widely used in game engines, physics simulations, and animation. While prior works can capture the global structure of input shapes, they may fail to preserve fine-grained details (e.g., filling a toaster's slots), which are critical for retaining the functionality of objects in interactive environments. In this paper, we propose a novel method that addresses the limitations of existing approaches from three perspectives: (a) We introduce a novel collision-aware concavity metric that examines the distance between a shape and its convex hull from both the boundary and the interior. The proposed concavity preserves collision conditions and is more robust to detect various approximation errors. (b) We decompose shapes by directly cutting meshes with 3D planes. It ensures generated convex hulls are intersection-free and avoids voxelization errors. (c) Instead of using a one-step greedy strategy, we propose employing a multi-step tree search to determine the cutting planes, which leads to a globally better solution and avoids unnecessary cuttings. Through extensive evaluation on a large-scale articulated object dataset, we show that our method generates decompositions closer to the original shape with fewer components. It thus supports delicate and efficient object interaction in downstream applications.
\end{abstract}



\begin{teaserfigure}
  \centering
  \includegraphics[width=\textwidth]{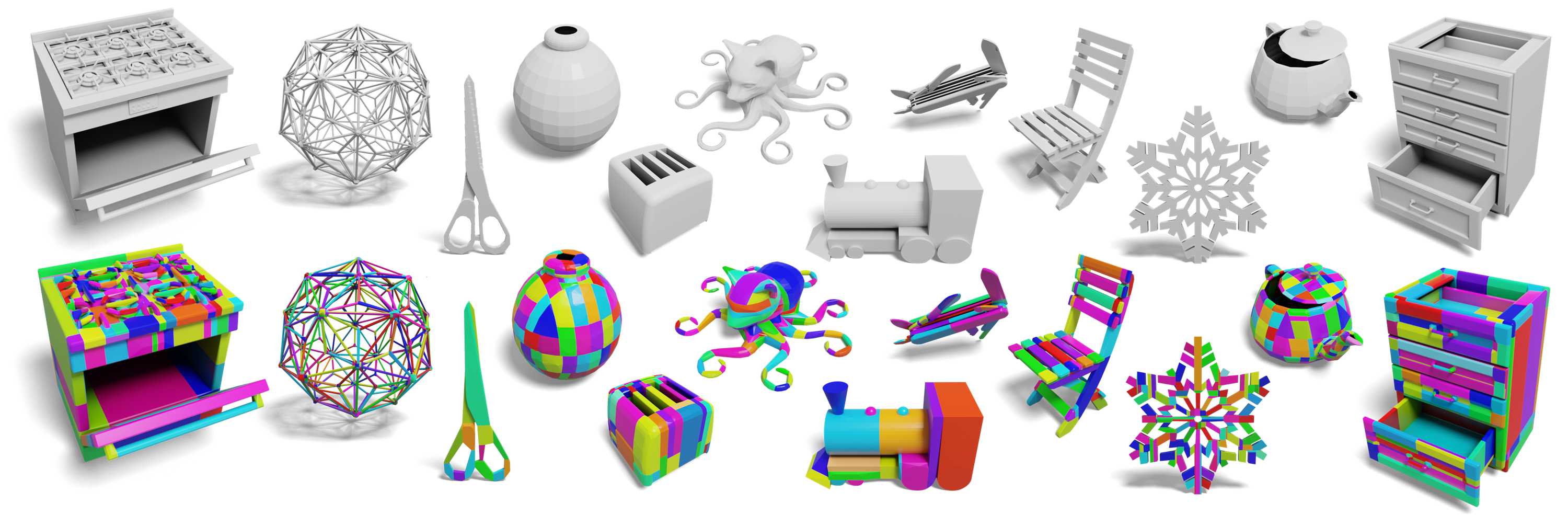}
  \caption{We decompose a solid mesh into a set of components and utilize the convex hulls of the components (shown in different colors) to represent the original shape. Compared to prior works, we can better capture the fine-grained structures of the input shape with fewer components. See handles of the oven and the cabinet, slots of the toaster, and the spout of the kettle (zoom in for details). The high-quality decomposition enables delicate object interaction in downstream applications (e.g., a robot opens the drawer by grabbing the handles).}
  \label{fig:teaser}
\end{teaserfigure}

\maketitle

\section{Introduction}

With the development of 3D depth sensors, VR/AR, and physics simulation, large-scale detailed 3D models have become more common. In addition to employing data structures such as octrees and bounding volume hierarchies (BVH) to accelerate specific geometry processing algorithms, another common strategy for handling complex 3D models is to decompose them into simpler components. In particular, convex decomposition has aroused great interest. Many fundamental geometry problems in rendering and physics simulation are non-trivial and computationally expensive to solve for general shapes. However, if input shapes are convex polyhedra, many of them can be formulated as convex optimization problems, and efficient algorithms can be specifically designed. Examples include determining whether a 3D point lies inside or outside of a mesh~\cite{snoeyink2017point}, checking whether two meshes intersect~\cite{liu2008convex, mirtich1998v,bergen1999fast}, and calculating the minimum distance between two meshes~\cite{gilbert1988fast}. 

Decomposing a 3D solid shape into a minimum number of exact convex components is the \textit{exact convex decomposition} (ECD) problem, which has proven to be NP-hard~\cite{o1983some,chazelle1997strategies}. Although many suboptimal heuristics~\cite{chazelle1981convex} have been proposed, they usually output a large number of small components, preventing them from practical applications. Instead, the \textit{approximate convex decomposition} (ACD) problem~\cite{lien2007approximate} proposes to lift the strict convexity constraint and only requires the decomposed components to be approximately convex. Since ACD approaches~\cite{lien2007approximate, mamou2009simple, mamou2016volumetric, thul2018approximate} typically generate a much smaller number of components, whose convex hulls can then be used to approximate the original shape and speed up downstream applications, ACD works have recently received 
more attention. For example, V-HACD~\cite{mamou2016volumetric} is currently one of the most popular open-source ACD methods and has been adopted by a wide range of game engines and physics simulation SDKs.

Existing ACD methods share a similar overall pipeline. In order to quantify the decomposition quality, they first define a concavity metric to measure the similarity between a decomposed component and its convex hull. They then design a heuristic cost function to decompose the 3D meshes greedily. There are three major shortcomings of existing ACD methods: \textbf{(a) Concavity metric:} Prior works mainly utilize two types of metrics: boundary-distance-based concavity~\cite{lien2004approximate, lien2007approximate, lien2008approximate,ghosh2013fast,liu2016nearly,mamou2009simple}, which measures the distance between the boundary surfaces of the shape and its convex hull; and volume-based concavity~\cite{mamou2016volumetric,attene2008hierarchical,thul2018approximate}, which calculates the volume difference between the solid shape and its convex hull. However, both metrics may fail to preserve the collision conditions in some cases, which means some positions in the space are unlikely to collide shape, but collide with the decomposition results. The insensitivity of existing concavity metrics to changes in collision conditions can be fatal for preserving object functionality. For example, they might cause an algorithm to stuff the slots of a toaster.  \textbf{(b) Component representation:} There are two common strategies for representing components and decomposing shapes. The first one is to decompose shapes by grouping the triangle faces~\cite{mamou2009simple,lien2007approximate,liu2016nearly}, which results in zig-zag boundaries of the components and intersecting convex hulls. In contrast, V-HACD~\cite{mamou2016volumetric} first voxelizes the input mesh and then decomposes the voxels. However, the voxelization introduces discretization artifacts, which even makes the algorithm unable to recognize already convex shapes. \textbf{(c) One-step greedy search:} Most previous works~\cite{mamou2016volumetric,thul2018approximate} decompose the shapes by recursively performing locally optimal actions. They often take short-sighted actions and end up generating more components. Furthermore, considering only one step may lead to various corner cases, which requires different heuristic terms~\cite{mamou2016volumetric} as workarounds. 

In this paper, we introduce a novel approximate convex decomposition method for 3D meshes, which effectively addresses the limitations of existing approaches from three corresponding perspectives: (a) We propose a novel collision-aware concavity metric that examines the component from both the boundary surface and shape interior by sampling points and calculating Hausdorff distance. The proposed concavity encourages preserving the collision conditions by penalizing the inclusion of regions that are far away from the original shape. We also propose an efficient way to calculate the concavity to speed up the decomposition. (b) We decompose shapes by directly cutting 3D solid meshes with 3D planes, which results in flat boundaries between components. It ensures intersection-free convex hulls and avoids the defects caused by voxelization. We also provide a lightweight mesh cutting implementation, which is 100x faster than off-the-shelf libraries. (c) We propose utilizing the Monte Carlo tree search to determine cutting planes, which simulates and searches multiple future actions before each cutting. Compared to the one-step greedy search, we are more likely to find cutting planes that lead to a better global solution and avoid unnecessary cuttings. In addition, by considering multiple steps, we no longer need various heuristic terms to prevent corner cases.

We evaluate our method on the V-HACD benchmark~\cite{mamou2016volumetric} and PartNet Mobility~\cite{xiang2020sapien}, a large-scale articulated object dataset. We show that our method better preserves the collision conditions and accurately approximates the fine-grained structures (e.g., drawer handles, kettle spouts, inner rings of scissors, and toaster slots) with fewer convex components. Our decomposition results thus enable delicate and fast object interaction in downstream applications. See Figure~\ref{fig:teaser} for some examples. Our code is available at \url{https://github.com/SarahWeiii/CoACD}. 

\section{Related Work}

\subsection{Application of Convex Shapes}
Many efficient geometric algorithms require convex shapes as input. For example, collision detection between shapes is the cornerstone in physics simulation, virtual reality, game engines, and animations. Fast and precise collision detection algorithms have been specially designed for convex shapes~\cite{liu2008convex, mirtich1998v, gilbert1988fast,bergen1999fast, weller2013brief}. Point location, which checks if a point is within a shape, is an important task in rendering and simulation. It can also be accelerated if input shapes are convex~\cite{snoeyink2017point}. Moreover, by abstracting a 3D shape with a set of convex components, many downstream applications are developed, such as skeleton extraction~\cite{lien2006simultaneous}, tetrahedral mesh generation~\cite{joe1994tetrahedral}, mesh deformation~\cite{xian2012automatic,wicke2007finite,jacobson2011bounded,wang2015linear,liu2021deepmetahandles}, and real-time animation~\cite{muller2013real}.

\subsection{Convex Decomposition}
\label{sec:related_work_convxe_decomposition}

The problem of \textit{exact convex decomposition} (ECD) was proved to be NP-hard~\cite{o1983some,chazelle1997strategies}, and many suboptimal heuristic algorithms have been proposed~\cite{chazelle1981convex,chazelle1984convex,bajaj1992convex, joe1994tetrahedral, bajaj1996splitting, hershberger1998erased}. However, ECD works typically outputs a substantial amount of components and slows down practical applications. People thus turned to \textit{approximate convex decomposition} (ACD)~\cite{lien2007approximate,lien2004approximate}, which lifts the strict convexity constraint and only requires the decomposed components to be almost convex.

ACD works first define a concavity metric, which measures the difference between a shape and its convex hull. They then iteratively decompose a 3D shape through top-down partition or bottom-up clustering, until the concavity of each decomposed component is within a pre-defined threshold. There are mainly three families of concavity metrics: 

\noindent\textbf{Boundary-distance-based:} \citet{lien2004approximate, lien2007approximate, lien2008approximate} propose to measure the distance between the shape boundary and its convex hull by mimicking the process of inflating a balloon. FACD~\cite{ghosh2013fast} further extends this idea by introducing a relative concavity to enhance the details of local structures. CoRise\cite{liu2016nearly} utilizes the shortest geodesic paths on the mesh surface and calculates the distance between the points on the path and edges of the convex hull. HACD~\cite{mamou2009simple} projects the mesh vertices to the convex hull surface along normal directions and then measures the distance between the vertices and their projection. Most boundary-distance-based concavities involve intricate geometric processing and are inefficient to calculate for 3D shapes. Moreover, only requiring a small distance between the shape boundary and its convex hull is not enough to guarantee plausible decomposition. 

\noindent\textbf{Volume-based:} There are also many works utilizing the volume difference between the shape and its convex hull as the concavity metric. \citet{attene2008hierarchical} tetrahedralizes the input mesh and hierarchically cluster the tets using the volume-based concavity. V-HACD~\cite{mamou2016volumetric} first voxelizes the input mesh and then greedily decomposes the voxels with axis-aligned cutting planes and volume-based concavity. Due to its open-source code and good performance in general cases, V-HACD is currently one of the most widely used convex decomposition algorithms. \citet{thul2018approximate} also utilizes the volume-based metric and extends the task from a single static mesh to temporal coherent animated meshes. While computationally efficient, many unreasonable decompositions may not be penalized using volume differences alone. For example, it's easy for V-HACD to stuff the slots of a toaster since the relative volume difference may be small.

\noindent\textbf{Visibility-based:} \citet{liu2010convex,ren2011minimum} count the pairs of surface points that are mutually visible within the inner volume of the shape, and utilize the ratio of visible pairs as the concavity metric. However, it may be biased by the positions of the concave parts and inefficient to calculate for complex shapes.

Recently, many learning-based methods~\cite{deng2020cvxnet, chen2020bsp} also attempted to represent 3D meshes with assemblies of convex polyhedra. However, they generate convex components based on a global embedding feature of the shape and may thus fail to preserve many input structures. Their poor generalizability on novel shapes also prevents them from extensive practical use.

\subsection{Shape Abstraction}

Another related direction is shape abstraction. Unlike convex decomposition, which accurately approximates original shapes, shape abstraction extracts the global structure and pays less attention to low-level details. Existing approaches utilize conventional optimization or deep learning to abstract a 3D shape into a set of primitives, such as cuboids~\cite{tulsiani2017learning,sun2019learning,gadelha2020learning,zou20173d,mo2019structurenet,smirnov2020deep,calderon2017bounding}, superquadrics~\cite{paschalidou2019superquadrics,paschalidou2020learning}, sphere-trees~\cite{bradshaw2004adaptive}, sphere-meshes ~\cite{gadelha2020learning,thiery2013sphere}, generalized cylinders~\cite{zhou2015generalized}, and their combination~\cite{li2019supervised}. Although most geometric primitives are convex, shape abstraction is hard to capture fine-grained structures due to the low-dimensional expressibility of the primitives.

\section{Problem Definition and Method Overview}
\label{sec:overview}
We aim to decompose a solid shape $\mathcal{S}$ (represented by a 2-manifold mesh) into a set of almost convex polyhedra $\{\mathcal{S}_i\}$, such that the solid shape determined by the union $\bigcup \left\{\mathcal{S}_{i} \right\}$ is equivalent to the original shape $\mathcal{S}$, and there is no intersection between the components $\mathcal{S}_{i}$ except for the boundary. We utilize a \textit{concavity} metric to measure the difference between each component and its convex hull. Our objective is to minimize the number of components while ensuring the concavity of each component is within a pre-defined threshold $\epsilon$. After decomposition, the convex hulls of the generated components can be used to approximate the original shape $\mathcal{S}$. By adjusting the threshold $\epsilon$, users can balance the number of components and the level of detail of the decomposition. We assume that the input is a 2-manifold solid mesh and we can convert imperfect input (e.g., non-watertight or non-manifold meshes) by pre-processing with an off-the-shelf manifold conversion algorithm~\cite{huang2018robust}.

\begin{algorithm}[t]
\DontPrintSemicolon
  \KwInput{A 2-manifold solid mesh $\mathcal{S}$, a concavity threshold $\epsilon$}
  \KwOutput{Approximate convex decomposition $\mathcal{D}$}
  $\mathcal{Q} \leftarrow\{\mathcal{S}\}$ \tcp*{processing queue}
  $\mathcal{D} \leftarrow \emptyset$ \tcp*{decomposition results}
  \While{$\mathcal{Q}$ is not empty}
   {
   		$C \leftarrow Q.dequeue()$ \;
   		\If (\tcp*[f]{Section~\ref{sec:concavity}}){$\widetilde{\operatorname{Concavity}}(C) < \epsilon$}
        {
            $\mathcal{D}.enqueue(C)$ \;
        }
        \Else
        {
        	$\mathcal{P} \leftarrow \operatorname{MCTS}(C)$   \tcp*{search for cutting plane, Section~\ref{sec:mcts}} \label{line:cut0}
        	$\mathcal{P} \leftarrow \operatorname{Refine}(C, \mathcal{P})$ \tcp*{Section~\ref{sec:refinement}} \label{line:cut1}
        	$C_L, C_R \leftarrow \operatorname{Cut}(C, \mathcal{P})$ \tcp*{Section~\ref{sec:cut}}
        	\label{line:cut2}
        	$\mathcal{Q}.enqueue(C_L,C_R)$ \;
        }
   }
   $\mathcal{D} \leftarrow \operatorname{Merge}(\mathcal{D, \epsilon})$ \tcp*{Section~\ref{sec:refinement}} \label{line:merge}
\caption{Approximate Convex Decomposition}
\label{alg:acd}
\end{algorithm}

As shown in Algorithm~\ref{alg:acd}, we utilize a divide-and-conquer strategy to recursively decompose the solid shape. For each component whose concavity is greater than the threshold $\epsilon$, we search for the best cutting plane and use it to split the component into two (\cref{line:cut0,line:cut1,line:cut2}). The cutting process is recursively applied until all the components satisfy the concavity constraint. At last, a post-processing step is applied in order to merge the generated components and further reduce the number of components (\cref{line:merge}).

Our high-level pipeline is similar to V-HACD, and the differences mainly come from three aspects. First, we introduce a novel collision-aware concavity metric that is more sensitive to detect various approximation errors and leads to a more reasonable and robust decomposition. Also, we propose an efficient way to calculate the concavity (Section~\ref{sec:concavity}). Second, without voxelizing the input mesh and splitting the voxels, we directly decompose the shapes by cutting meshes with 3D planes, which avoids the discretization errors and supports more precise and efficient decomposition (Section~\ref{sec:cut}). Third, instead of greedily searching for the locally optimal decompose actions with one-step results, we propose leveraging multi-step tree search to achieve globally better decomposition (Section~\ref{sec:mcts}).  

\section{Collision-Aware Concavity Metric}
\label{sec:concavity}

\subsection{Concavity Definition}

ACD works utilize \textit{concavity} to measure the difference between a solid shape $\mathcal{S}$ and its convex hull $\operatorname{CH}(\mathcal{S})$, which can be used to quantify the quality of the decomposed components. The ideal \textit{concavity} should be able to recognize all unreasonable approximations and penalize them with a high cost. It should also be efficient to calculate, as numerous concavity calculations are needed during the decomposition process. 

\begin{figure}[t]
\begin{center}
   \includegraphics[width=\linewidth]{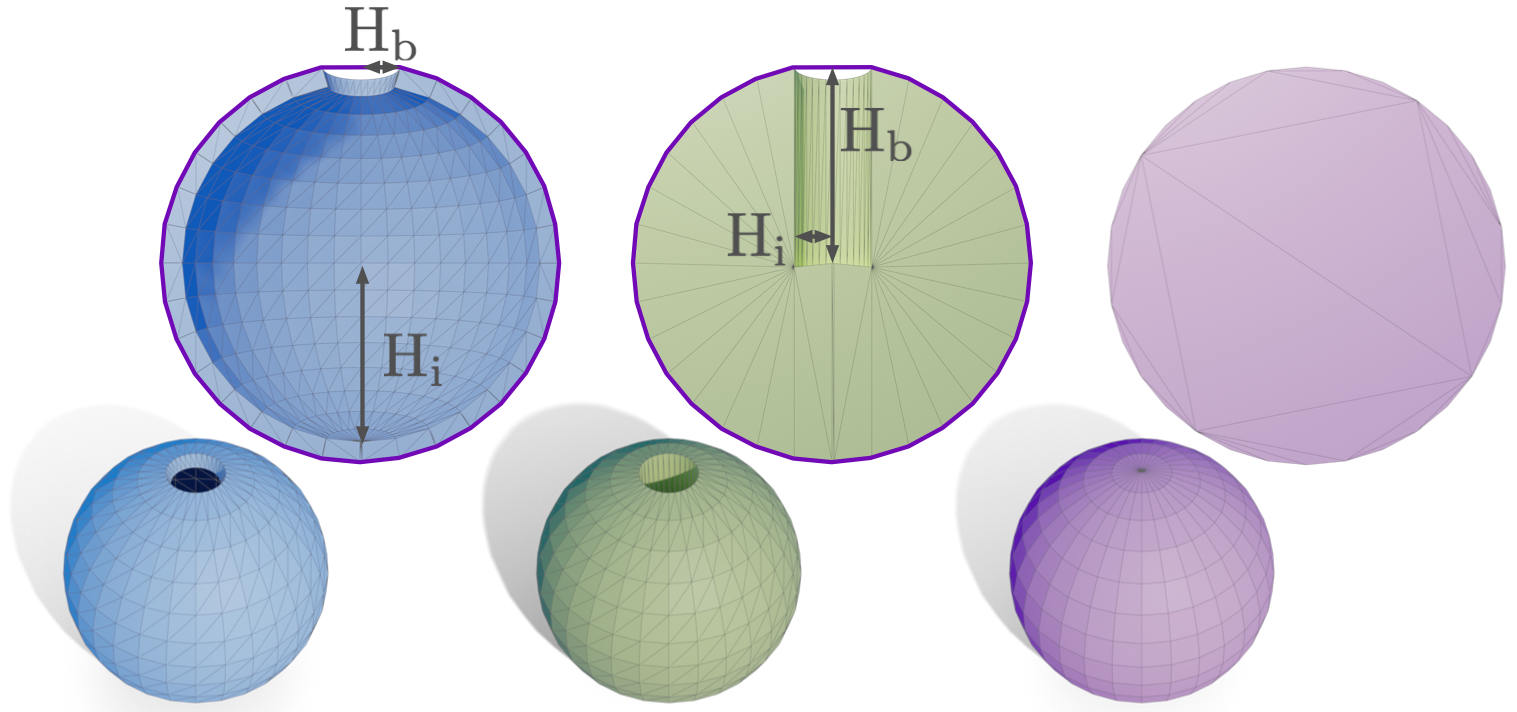}
\end{center}
  
  \caption{From left to right: a spherical shell with a small opening, a solid sphere with a deep hole, and a solid sphere indicating the convex hull of the left two shapes. Both $\operatorname{H_i}(\mathcal{S})$ and $\operatorname{H_b}(\mathcal{S})$ are necessary to measure the difference between a shape and its convex hull. In the blue example, $\operatorname{H_i}(\mathcal{S})\gg\operatorname{H_b}(\mathcal{S})$, while in the green example, $\operatorname{H_b}(\mathcal{S})\gg\operatorname{H_i}(\mathcal{S})$. The purple polygons surrounding the cross-sections indicate the boundary surface of the convex hull.}
\label{fig:hi_hb}
\end{figure}

There is no consensus on the definition of the concavity among existing ACD works. Some of them~\cite{lien2004approximate, lien2007approximate, lien2008approximate,ghosh2013fast,liu2016nearly,mamou2009simple} focus on the distance between the boundary surfaces of $\mathcal{S}$ and $\operatorname{CH}(\mathcal{S})$, while other works~\cite{mamou2016volumetric,attene2008hierarchical,thul2018approximate} utilize the volume difference between $\mathcal{S}$ and $\operatorname{CH}(\mathcal{S})$ as the concavity. Please refer to Section~\ref{sec:related_work_convxe_decomposition} for more details. 

A reasonable decomposition should preserve the collision conditions of the input shape, which means any position in the space that is unlikely to collide with the input shape (i.e., far away from the input shape), should not collide with the convex decomposition results as well. Otherwise, both the structure and functionality of the input shape can be greatly affected. A good concavity metric should thus be sensitive to detect the approximation errors that significantly change the collision conditions. However, we argue that both the boundary-distance-based and the volume-based metrics may fail to do so and lead to undesirable decomposition results: 

\begin{figure}[t]
\begin{center}
   \includegraphics[width=\linewidth]{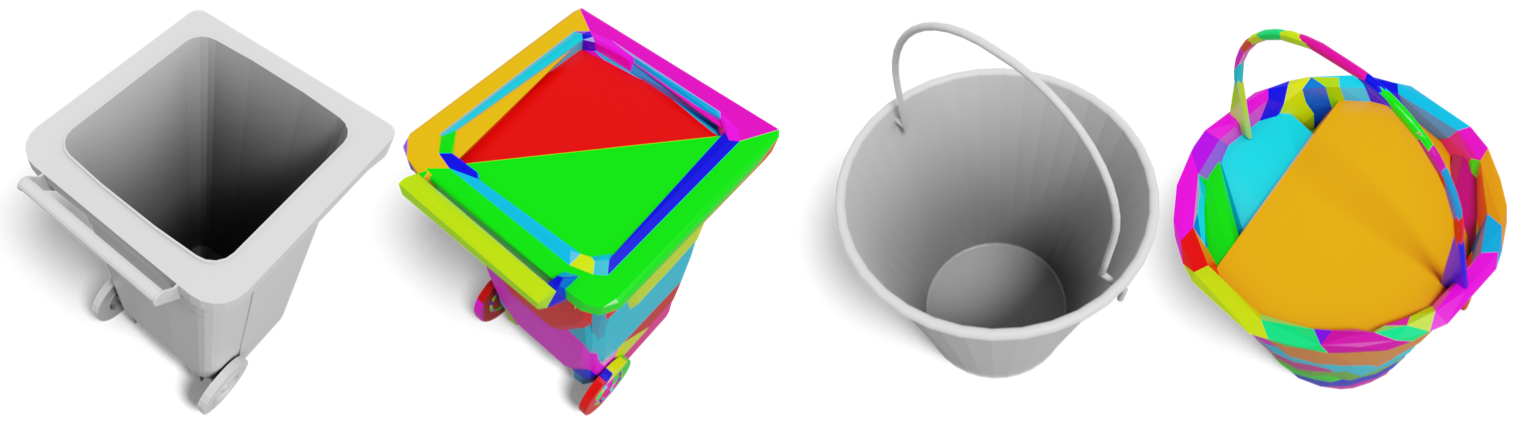}
\end{center}
  \caption{Failure cases of the boundary-distance-based methods (from HACD~\cite{mamou2009simple}). Focusing only on the boundary distance between the shape and its convex hull, HACD may fail to handle the hollow structures and fill the interior space.}
\label{fig:hacd_failure}
\end{figure}

\begin{figure}[t]
\begin{center}
    \centering
   \includegraphics[width=\linewidth]{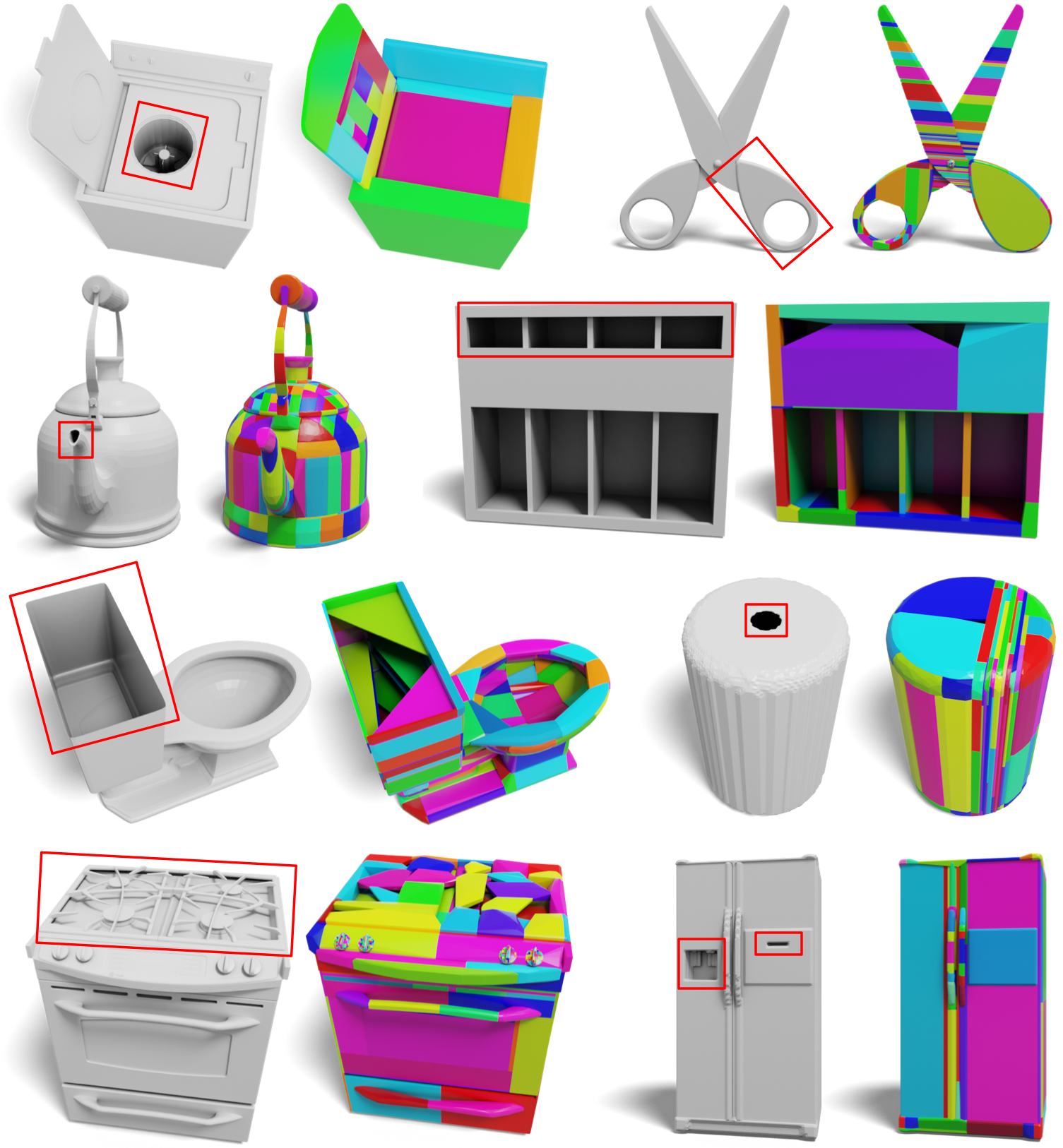}
\end{center}
  \caption{Failure cases of the volume-based methods (from V-HACD~\cite{mamou2016volumetric}). Focusing on the volume difference, V-HACD may fill holes when the relative volume of the errors is not too large (e.g., thin planar structures). The red rectangles highlight the error-prone regions. }
\label{fig:vhacd_failure}
\end{figure}

\begin{itemize}[leftmargin=*]
    \item \emph{Boundary-distance-based metrics alone are insufficient for preserving collision conditions.} As shown in Figure~\ref{fig:hi_hb}, if $\mathcal{S}$ is a hollow spherical shell (blue example), the boundary distance between $\mathcal{S}$ and $\operatorname{CH}(\mathcal{S})$ may be quite small, and the algorithm may thus fill the interior space and approximate $\mathcal{S}$ with a solid convex hull. However, it is inappropriate to do so if we want to exploit the free space inside $\mathcal{S}$. In fact, such shell-like structures are quite common in applications. For example, in physical simulators, a teapot needs to hold water, and it's inappropriate to approximate the body of the teapot with a solid convex hull. Otherwise, the water particles collide with the interior of the teapot. Figure~\ref{fig:hacd_failure} shows some failure cases of the boundary-distance-based concavity.
    \item \emph{Volume-based metrics alone are insufficient for preserving collision conditions.} In some cases, the volume difference between $\mathcal{S}$ and $\operatorname{CH}(\mathcal{S})$ may be very small, but significant differences may exist at the boundary of $\mathcal{S}$ and $\operatorname{CH}(\mathcal{S})$. For example, in Figure~\ref{fig:hi_hb}, if $\mathcal{S}$ is a solid sphere with a deep hole (green one), the relative volume of the deep hole is small. However, it is inappropriate to approximate $\mathcal{S}$ with its convex hull, which fills the hole. Figure~\ref{fig:vhacd_failure} shows common failure cases of the volume-based concavity, where V-HACD tends to utilize thin planar components to close the holes.
\end{itemize}

\begin{figure}[t]
\begin{center}
   \includegraphics[width=\linewidth]{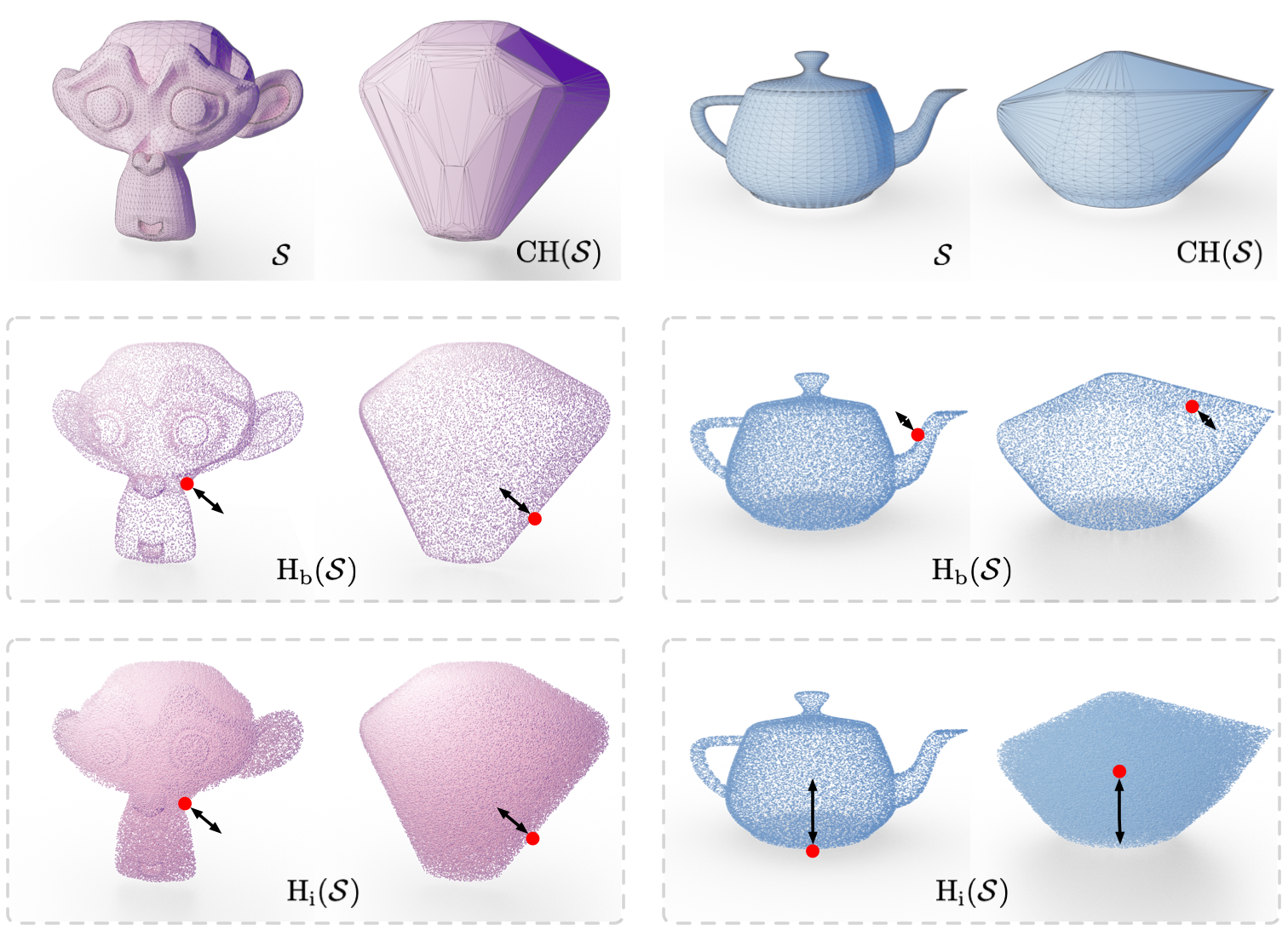}
\end{center}
  \caption{The figure illustrates how to calculate the concavity for ``Blender's Suzanne'' and ``Utah teapot''. The first and third columns show the input shapes and their sampled point clouds, while the second and fourth columns show the convex hulls and their sampled point clouds. From top to bottom: manifold meshes, point clouds sampled from the boundary surfaces (for calculating $\operatorname{H_b}(\mathcal{S})$), and point clouds sampled from the interior of the shapes (for calculating $\operatorname{H_i}(\mathcal{S})$). The teapot is hollow since its inside is connected to the outside world through the spout. In each dotted square, the red points indicate the pair of points that achieves $\operatorname{H_i}(\mathcal{S})$ or $\operatorname{H_b}(\mathcal{S})$. }
 \label{fig:concavity_definition}
\end{figure}

The failure cases of existing concavity metrics are fatal to the functionality of the objects in downstream applications. For example, suppose we use the decomposition results shown in Figure~\ref{fig:vhacd_failure} in a simulator. In that case, robots can no longer grasp the scissors by the circles, can no longer use the kettle to pour water, and can no longer interact with the water dispenser on the refrigerator. 

However, it's non-trivial to directly combine the existing boundary-distance-based metrics and volume-based metrics since they have different geometric meanings and scales. On the other hand, many existing boundary-distance-based metrics involve intricate geometry processing and are cumbersome to calculate.

To overcome the limitations of existing approaches, we propose a novel \textit{collision-aware concavity metric} that examines the decomposed component from both the boundary and the interior with a unified metric. To introduce the metric, we first review the definition of the Hausdorff distance for two point sets: 
\begin{equation}
    \operatorname{H}(A, B) = \max \{\sup_{a \in A}d(a, B), \sup_{b \in B}d(b, A)\}
\end{equation}
where $A$ and $B$ are two point sets, $d(x, Y) = \inf_{y \in Y}d(x, y)$ and $d(x, y)$ indicates the Euclidean distance between the two points.

As shown in Figure~\ref{fig:concavity_definition}, we sample two pairs of point sets to measure the distance between a solid shape $\mathcal{S}$ and its convex hull $\operatorname{CH}(\mathcal{S})$. Specifically, by sampling the points from the two boundary surfaces, we define $\operatorname{H_b}(\mathcal{S})$ as:
\begin{equation}
    \operatorname{H_b}(\mathcal{S}) = \operatorname{H}(\operatorname{Sample}(\partial \mathcal{S}), \operatorname{Sample}(\partial \operatorname{CH}(\mathcal{S})))
\end{equation}
where $\operatorname{Sample}()$ indicates the point set sampling operation, and $\partial$ denotes the boundary surface of a solid shape. Similarly, by sampling points from the interior of the shapes, we define $\operatorname{H_i}(\mathcal{S})$ as:
\begin{equation}
    \operatorname{H_i}(\mathcal{S}) = \operatorname{H}(\operatorname{Sample}(\operatorname{Int} \mathcal{S}), \operatorname{Sample}(\operatorname{Int} \operatorname{CH}(\mathcal{S})))
\end{equation}
where $\operatorname{Int}$ denotes the interior of a solid shape. 

The \textit{concavity} of a solid shape $\mathcal{S}$ is then proposed to be:
\begin{equation}
    \operatorname{Concavity}(\mathcal{S}) = \max (\operatorname{H_b}(\mathcal{S}), \operatorname{H_i}(\mathcal{S}))
    \label{equ:concavity}
\end{equation}
By measuring the distance from both the boundary surface and the interior, the proposed concavity can better capture shape differences and well address failure cases of the prior concavity metrics. 

We argue that both terms in Equation~\ref{equ:concavity}  are necessary and complementary for measuring the shape difference. Only using $\operatorname{H_b}(\mathcal{S})$ may fail to handle the shell-like structures. Taking the spherical shell (the blue one shown in Figure~\ref{fig:hi_hb}) as an example, $\operatorname{H_b}(\mathcal{S})$ only measures the radius of the small opening on the boundary surface, which is quite small. In contrast, $\operatorname{H_i}(\mathcal{S})$ recognizes the large difference from the interior and penalizes it with the inner radius of the shell. On the other hand, only using $\operatorname{H_i}(\mathcal{S})$ may fail to capture the difference between boundary surfaces. For the solid sphere with a deep hole (the green one shown in Figure~\ref{fig:hi_hb}), $\operatorname{H_i}(\mathcal{S})$ only measures the radius of the hole, no matter how deep the hole is. Instead, $\operatorname{H_b}(\mathcal{S})$ is able to measure the depth of the hole, which better captures the difference. 

A notable property of the proposed concavity metric is its \textit{collision awareness}.  Our concavity encourages decompositions to preserve the collision conditions by penalizing the distances between points in $\operatorname{CH}(\mathcal{S}) - \mathcal{S}$ (i.e., the extra volume introduced by the convex hull) and the original shape $\mathcal{S}$. Therefore, our metric is sensitive to detecting approximation errors that significantly alter the collision conditions, no matter they are fine-grained structures with small volume or thin planar structures. Instead of calculating an overall average difference, the proposed concavity focuses on the worst case (i.e., the farthest point pair). This is because that in many applications (e.g., robot simulation), we need a guarantee about the worst case to avoid fatal approximations to certain parts of the shape, even though the approximation may look good overall.

Moreover, by using our proposed metric, it's more intuitive for users to set and adjust the concavity threshold $\epsilon$, since one can interpret the threshold as the degree to which the original shape becomes thicker. In contrast, the volume-based concavity may not have such an intuitive interpretation, and the change caused by adjusting the volume difference threshold may be less predictable.

\subsection{Efficient Concavity Calculation}
Besides recognizing all implausible approximations, a good concavity metric should also be efficient to calculate. Our proposed metric samples points from the shape and its convex hull, and then calculates Hausdorff distance between the point sets, which avoids intricate geometry processing required by existing boundary-distance-based metrics. The nearest neighbor calculation in Hausdorff distance can be accelerated by approximation approaches and parallel computation~\cite{blanco2014nanoflann,arya1998optimal}. Moreover, when calculating $\operatorname{H_b}(\mathcal{S})$, we exploit point-to-triangle-face distances to improve the precision further and reduce the number of samples.

That being said, calculating accurate $\operatorname{H_i}(\mathcal{S})$ can still be time-consuming. The calculation of $\operatorname{H_b}(\mathcal{S})$ only needs sampled points from the boundary surfaces, while the calculation of $\operatorname{H_i}(\mathcal{S})$ needs much more sampled points from the interiors of the shapes to achieve high precision. See Figure~\ref{fig:concavity_definition} for differences between the sampled points. To further accelerate the concavity calculation, we propose a surrogate term $\operatorname{R_v}(\mathcal{S})$ for $\operatorname{H_i}(\mathcal{S})$:

\begin{equation}
    \operatorname{R_v}(\mathcal{S}) = \sqrt[3]{\frac{3(\operatorname{Vol}(\operatorname{CH}(\mathcal{S})) - \operatorname{Vol}(\mathcal{S}))}{4\pi}}
\end{equation}
where $\operatorname{Vol}(\operatorname{CH}(\mathcal{S}))$ - $\operatorname{Vol}(\mathcal{S})$ indicates the volume difference between the convex hull and the input shape. The geometric interpretation of $\operatorname{R_v}(\mathcal{S})$ is the radius of a sphere with volume $\operatorname{Vol}(\operatorname{CH}(\mathcal{S}))$ - $\operatorname{Vol}(\mathcal{S})$, which is potentially the largest inscribed sphere within the difference of $\operatorname{CH}(\mathcal{S})$ and $\mathcal{S}$. $\operatorname{R_v}(\mathcal{S})$ serves a similar role as $\operatorname{H_i}(\mathcal{S})$ to recognize the differences within the interior of solid shapes.  Moreover, we can actually prove a theoretical guarantee for $\operatorname{R_v}(\mathcal{S})$:
\begin{theorem} For every solid shape $\mathcal{S}$, we have:
$$\sqrt{2}\max(\operatorname{H_b}(\mathcal{S}), \operatorname{R_v}(\mathcal{S})) \geq \max(\operatorname{H_b}(\mathcal{S}), \operatorname{H_i}(\mathcal{S}))$$
\end{theorem}
The theorem indicates that we can use $\operatorname{H_b}(\mathcal{S})$ and $\operatorname{R_v}(\mathcal{S})$ to bound the proposed $\operatorname{Concavity}(\mathcal{S})$ and still be able to recognize any unreasonable approximations. Please refer to the supplementary materials for detailed proof. In practice, we find that $\operatorname{R_v}(\mathcal{S})$ often overestimates $\operatorname{H_i}(\mathcal{S})$. We thus utilize:
\begin{equation}
    \widetilde{\operatorname{Concavity}}(\mathcal{S}) = \max(\operatorname{H_b}(\mathcal{S}), k\operatorname{R_v}(\mathcal{S}))
\end{equation}
where $k$ is a coefficient less than 1, as the concavity metric to achieve a better approximation. It's much faster to calculate $ \widetilde{\operatorname{Concavity}}(\mathcal{S})$, since it avoids sampling points from the interior of the shapes and the corresponding nearest neighbor calculation. 

\section{Shape Decomposition by Cutting Meshes}
\label{sec:cut}

\begin{figure}[t]
\begin{center}
   \includegraphics[width=0.83\linewidth]{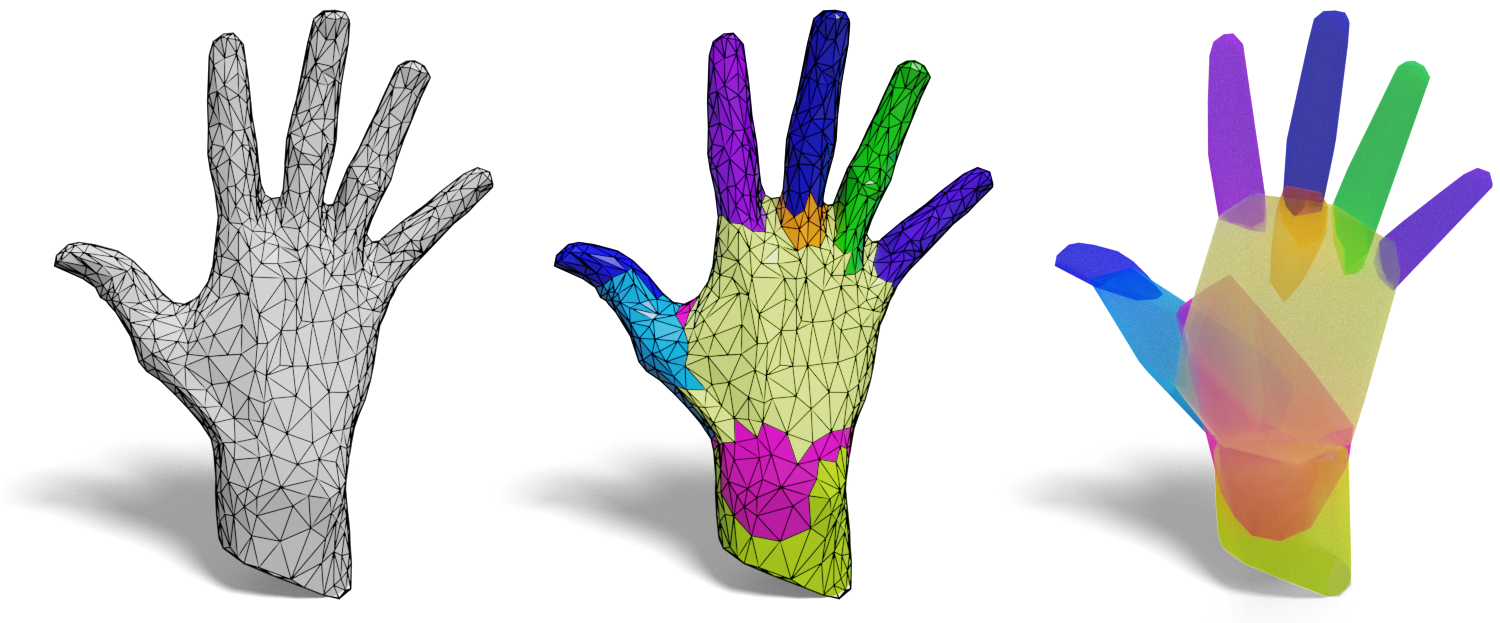}
\end{center}
  \caption{An example of triangle-grouping-based methods (from HACD). From left to right: (a) Input triangle mesh. (b) Grouping results of the triangle faces, where each color indicates a component. There are zig-zag boundaries between different components. (c) Corresponding convex hulls of each component, and they intersect with each other.}
  \label{fig:hand}
\end{figure}

\begin{figure}[t]
\begin{center}
   \includegraphics[width=\linewidth]{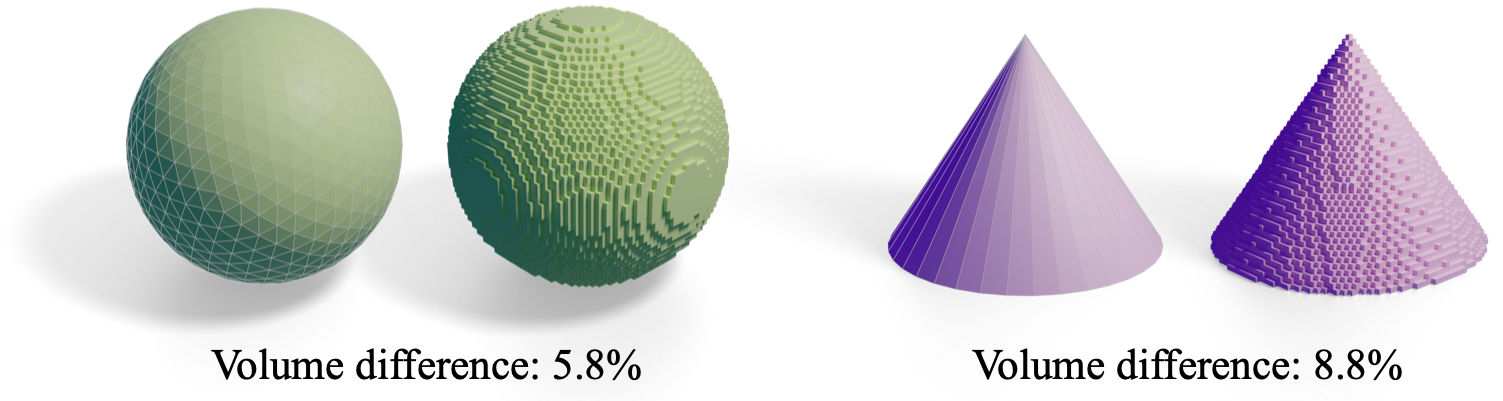}
\end{center}
  \caption{Each pair shows an input mesh and its voxelization ($64\times64\times64$). Even the input shapes are already convex, there are noticeable volume differences between the voxels and their bounding convex hulls, making V-HACD fail to recognize already convex parts.}
\label{fig:voxelization}
\end{figure}

As introduced in Section~\ref{sec:overview}, our method recursively decomposes a shape $\mathcal{S}$ into smaller components. There are various ways to represent a component and different geometry processing strategies to decompose the shape. We discuss the geometric design in this section and leave the search strategy to the next section.

In general, prior works mainly take two types of decomposing strategies. For triangle-grouping-based methods~\cite{mamou2009simple,liu2016nearly}, they preserve the triangle faces of the input mesh, and group the faces by top-down division or bottom-up clustering. For volume-based methods, like V-HACD, they first voxelize the input mesh, use voxels to represent the shape, and then divide the voxels. However, both strategies have some apparent limitations. Specifically, triangle-grouping-based methods often output components with zigzag boundaries (see Figure~\ref{fig:hand}). As a result, the convex hulls of the decomposed components usually intersect with each other, which is undesirable in many applications. On the other hand, although volume-based methods avoid crooked boundaries of the components, their voxelization pre-processing may introduce discretization artifacts. More importantly, volume-based methods may fail to recognize already convex components. As shown in Figure~\ref{fig:voxelization}, although the input shapes are already convex, V-HACD still regards them as non-convex due to the discretization error, and may try to further divide the voxels. Not only that, but to achieve high precision, V-HACD would require a large number of voxels which would slow down the decomposition algorithm.

\begin{wrapfigure}{l}{0.17\textwidth}
    \centering
    \includegraphics[width=0.2\textwidth]{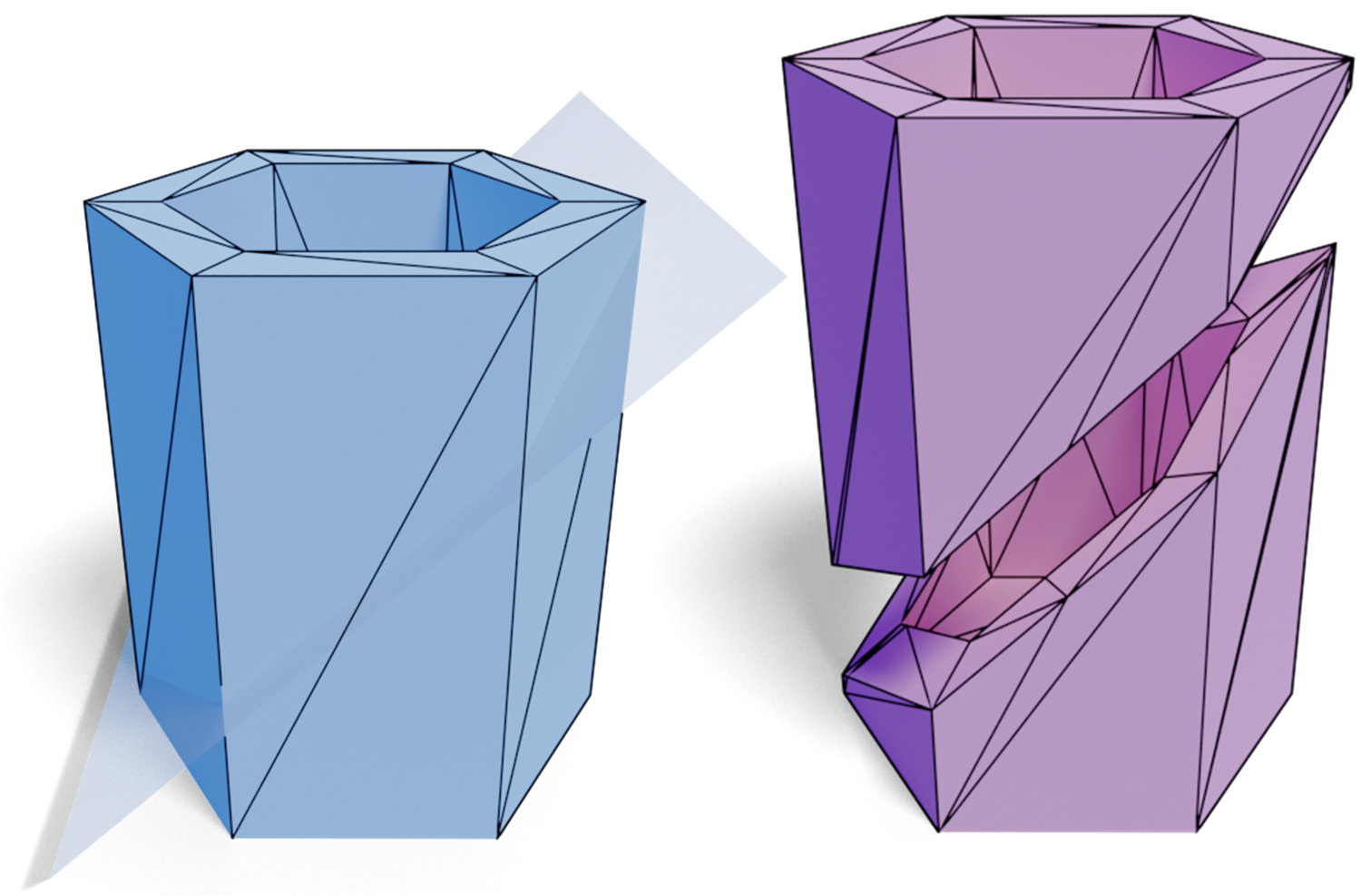}
\end{wrapfigure}
Instead, we follow~\cite{thul2018approximate} to utilize triangle meshes to represent solid components during decomposition and directly cut the manifold meshes with 3D planes. As shown in the left inset, given a manifold mesh and a 3D cutting plane, we split the mesh into two parts along the plane surface. Each resulting part is still a manifold mesh with flat boundaries and can be recursively decomposed. In this way, we ensure convex hulls of the decomposed components are intersection-free. Moreover, without voxelization as pre-preprocessing, we preserve fine-grained details and avoid over-decomposing convex shapes. 

However, existing mesh cutting functions from off-the-shelf computational geometry libraries (e.g., CGAL~\cite{fabri2009cgal}) are very heavy and time-consuming, which slows down our decomposition algorithm. Therefore, we implement a lightweight cutting function that is about 100x faster than CGAL's implementation. Specifically, the implementation mainly includes four steps: (a) Find triangles that are not intersecting with the cutting plane, and group them into two sets according to which side of the plane the triangle is in. The two triangle sets are later used to form the two parts. (b) Split each intersecting triangle into two with the cutting plane, and then add them into the two sets. (c) Add new surfaces (overlapping with the cutting plane) for the two parts to form solid meshes. To achieve this, we solve the constrained Delaunay triangulation~\cite{shewchuk1996triangle}, where the intersecting edges serve as boundary constraints. (d) Remove redundant triangles (if any) introduced in step (c), which correspond to holes on the newly added surfaces.  

\section{Monte Carlo Tree Search for Cutting Plane}
\begin{figure}[t]
\begin{center}
   \includegraphics[width=\linewidth]{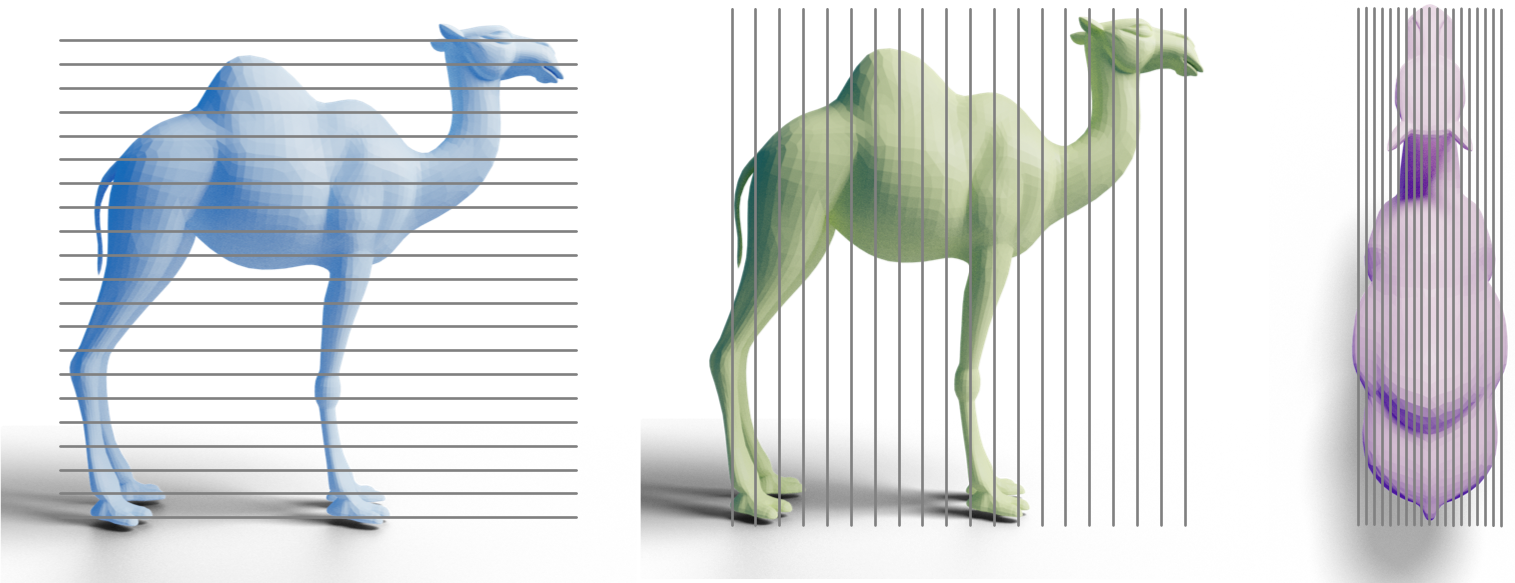}
\end{center}
  \caption{We sample $m$ equally-spaced candidate cutting planes (illustrated by straight lines here) from each axis-aligned direction. After finding the best candidate with a tree search, we also refine the plane's position.}
\label{fig:candidate_planes}
\end{figure}

\label{sec:mcts}
\subsection{Search Space}
During the decomposition process, for each intermediate component $C$ whose $\widetilde{\operatorname{Concavity}}(C)$ is greater than the threshold $\epsilon$, we find a cutting plane to split $C$ into two parts $C_L$ and $C_R$. Since there are infinite cutting planes in the 3D space, we follow V-HACD~\cite{mamou2016volumetric} to restrict the candidate planes to be axis-aligned (parallel to $xy$, $xz$, or $yz$ planes), and sample $m$ equal-spaced candidates along each direction, as shown in Figure~\ref{fig:candidate_planes}.  Axis-aligned discretized candidate planes enable a feasible search space and avoid irregular cutting results. The found optimal discretized candidate will be refined in the continuous local neighborhood for a more accurate cutting. (\cref{sec:refinement}) Also, we optionally perform a PCA~\cite{pearson1901liii} for the input shape at the beginning of the decomposition algorithm to align the cutting plane directions with the principal axes of the input shape.

\begin{figure}[t]
\begin{center}
   \includegraphics[width=\linewidth]{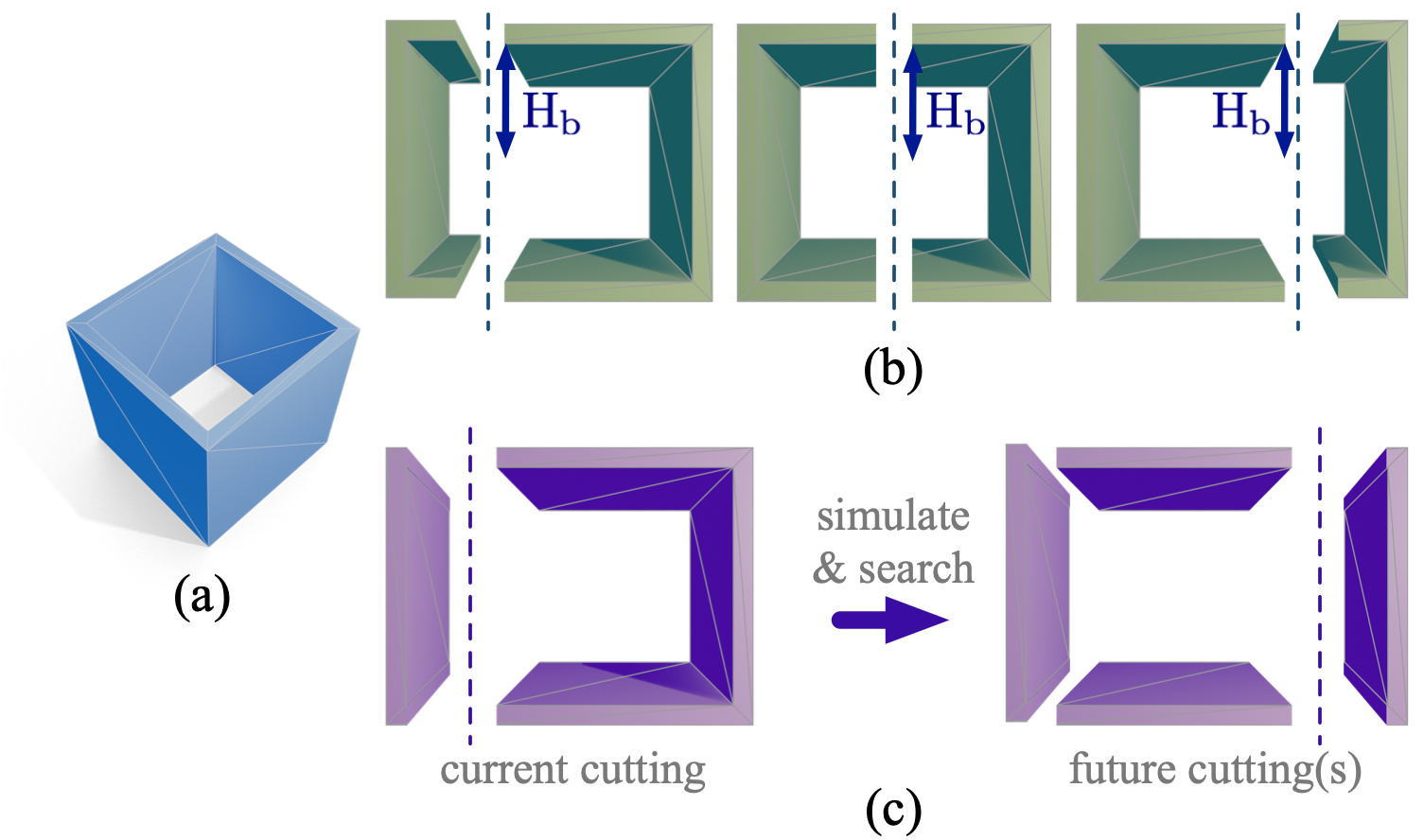}
\end{center}
  \caption{Comparison between one-step greedy and multi-step search. (a) Input shape (a cube without top and bottom). (b) The one-step greedy algorithm fails to find the proper first cutting plane, since all candidate cutting planes lead to the same cost (Equation~\ref{equ:concavity_deduction}) as illustrated by the blue arrows ($\operatorname{H_b}$). (c) The multi-step search algorithm can instead find the proper first cutting plane by simulating and searching future cuttings, which leads to the globally optimal solution (decomposed into exactly four pieces).}
 \label{fig:mcts_greedy}
\end{figure}

\begin{figure}[t]
\begin{center}
   \includegraphics[width=\linewidth]{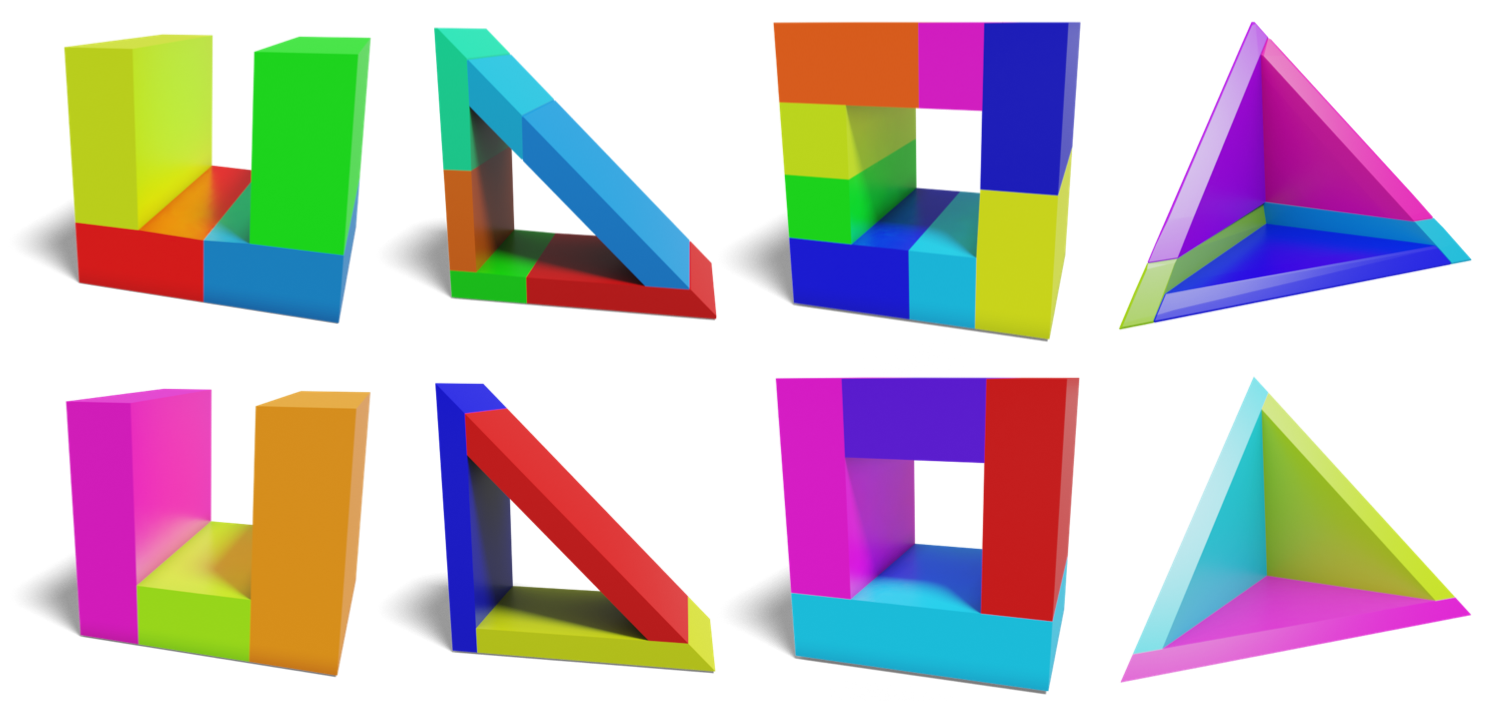}
\end{center}
  \caption{Failure cases of one-step greedy search. First row:  V-HACD employs a greedy search and generates redundant components. Second row: our method utilizes a multi-step tree search and solves the cases perfectly.}
\label{fig:vhacd_greedy}
\end{figure}

\subsection{One-Step Greedy vs. Multi-Step Search}
Intuitively, prior works~\cite{mamou2016volumetric,thul2018approximate} greedily find a cutting plane from the candidates, which minimizes:
\begin{equation}
    \max(\widetilde{\operatorname{Concavity}}(C_L), \widetilde{\operatorname{Concavity}}(C_R))
    \label{equ:concavity_deduction}
\end{equation} 
However, the one-step greedy search may be short-sighted and fail to find cutting planes that result in better global decomposition, and end up with more decomposed components. Figure \ref{fig:mcts_greedy} shows such an example, where the optimal solution is to decompose the input shape into exactly four square parts. However, when only one step is taken into account, the greedy search tries to cut the shape from the middle, resulting in more components. Figure~\ref{fig:vhacd_greedy} compares decomposition results of some simple primitives. Since V-HACD utilizes a greedy search, it fails to decompose the cases perfectly. 

Moreover, for the one-step greedy search, only using a concavity metric to find the cutting plane may often produce poor results and even block the decomposition algorithm. For example, when only considering one cutting, many candidate planes may lead to the same concavity deduction (Equation~\ref{equ:concavity_deduction}), and the cutting plane selection thus becomes arbitrary in these draw situations. To this end, prior works~\cite{mamou2016volumetric} introduce various auxiliary heuristic terms (e.g., balance term and symmetry term) for Equation~\ref{equ:concavity_deduction} as a workaround to make the algorithm more robust to various cases.

\begin{algorithm}[t]
\DontPrintSemicolon
\SetKwFunction{FMCTS}{MCTS}
\SetKwFunction{FTREEPOLICY}{TreePolicy}
\SetKwFunction{FDEFAULTPOLICY}{DefaultPolicy}
\SetKwFunction{FQUALITY}{Quality}
\SetKwFunction{FBACKUP}{Backup}
\SetKwProg{Fn}{Function}{:}{}
  \Fn{\FMCTS{$C, t, d$}}{
        Create root node $v_0$ with input mesh $C$\;
        \While{within $t$ iterations}    
        { 
        	$\{\mathcal{P}_1, \cdots, \mathcal{P}_l\}, v_l\leftarrow $ \FTREEPOLICY($v_0$ $,d$) \;
        	$\{\mathcal{P}_{l+1}, \cdots, \mathcal{P}_{d}\} \leftarrow$ \FDEFAULTPOLICY($v_l$ $,d$) \;
        	$q \leftarrow$ \FQUALITY{$\{\mathcal{P}_{1}, \cdots, \mathcal{P}_{d}\}$} \tcp*{Calculate a score} \label{line:score}
        	\FBACKUP{$v_l$, $q$} \tcp*{Update the score along the tree path} \label{line:update}
        }
        $v^* \leftarrow \argmax\limits_{v^{\prime} \in \text{children of } v_0} Q(v^{\prime})$ \; \label{line:best_child}
        \KwRet corresponding plane of $v^*$\;
  } 
  
  \Fn{\FTREEPOLICY{$v$ $,d$}}{
    $\mathcal{S} \leftarrow \emptyset$ \tcp*{Selected cutting planes}
    \While (\tcp*[f]{From the root to the leaf}) {$depth(v) < d$} {
        $c^* = \argmax\limits_{c_i \in \text{components of } v} \operatorname{Concavity}(c_i)$ \;
        
        \If {all cutting plane candidates of $c^*$ are expanded} {
            $v \leftarrow$ best child of $v$ according to the UCB formula\; \label{line:UCT}
            $\mathcal{S} \leftarrow \mathcal{S} + \{ \text{corresponding plane of } v\}$
        }
        \Else (\tcp*[f]{Expand a new child for $v$})
        {
            Randomly select a untried cutting plane $\mathcal{P}$ of $c^*$\; \label{line:expand1}
            Cut $c^*$ into $c^*_l$ and $c^*_r$ with $\mathcal{P}$\; \label{line:expand2}
            Create a new child $v^{\prime}$ to $v$ with $\mathcal{P}, c^*_l$ and $c^*_r$ \; \label{line:expand3}
            \KwRet $\mathcal{S} + \{ \mathcal{P}\}, v^{\prime}$ \;
        }
    }
    \KwRet $\mathcal{S}, v$ \;
  }
  
  \Fn{\FDEFAULTPOLICY{$v$ $,d$}}{
    $\mathcal{S} \leftarrow \emptyset$ \tcp*{Selected cutting planes}
    $\{c_i\}  = \operatorname{Copy}(\text{components of }v)$  \tcp*{Avoid affecting tree nodes}
    \For{$i \in \operatorname{range}(d - depth(v))$} {
        $c^* = \argmax\limits_{c_i} \operatorname{Concavity}(c_i)$ \;
        \For {direction in $\{xy, xz, yz\}$ } {
            Try to cut $c^*$ into $c^*_l$ and $c^*_r$ from middle with a plane along the $direction$ \; \label{line:default_cut}
            $q \leftarrow -\max(\operatorname{Concavity}(c^*_l), \operatorname{Concavity}(c^*_r))$ \;
        }
        $\mathcal{P} \leftarrow $ cutting plane that lead to the largest $q$ \; 
        Cut $c^*$ into $c^*_l$ and $c^*_r$ with $\mathcal{P}$\;
        $\mathcal{S} \leftarrow \mathcal{S} + \{ \mathcal{P}\}$ \;
    }
    \KwRet $\mathcal{S}$ \;
  }
  \Fn{\FBACKUP{$v, q$}}{
    \While (\tcp*[f]{From the leaf to the root}) {$v$ is not null} {  
        $N(v) \leftarrow N(v) + 1$  \tcp*{Visit times} 
        $Q(v) \leftarrow \max(Q(v), q)$ \tcp*{Value function}  
        $v \leftarrow$ parent of $v$ \;
    }
  }
\caption{Search for Cutting Plane}
\label{alg:mcts}
\end{algorithm}

We instead propose to take multiple steps into account when searching for a cutting plane. Specifically, we solve a Monte Carlo tree search (MCTS), which simulates multiple future cuttings, to find a cutting plane for each intermediate component $C$. This way, we pay more attention to long-term interests and are more likely to find cutting planes that lead to global optimal decompositions. Moreover, we find that by considering multiple steps during the tree search, we no longer need any other heuristic terms to prevent various corner cases. 
\subsection{Search Tree Structure}
It's non-trivial to apply MCTS to our cutting plane search, and many dedicated designs are involved. Specifically, in our tree search, each node represents a set of decomposed components $\{c_i\}$ for $C$, and the root node contains a single component $\{C\}$. For each node, we aim to cut the component $c^*$ with the largest concavity among those components associated with the node, and each child node corresponds to a cutting action for $c^*$. Since a cutting plane splits $c^*$ into two parts, the number of a child node's components is equal to the number of its parent node's components plus one. Also, since we sample $m$ candidate cutting planes from each axis-aligned direction, each node contains at most $3m$ child nodes. 

As shown in Algorithm~\ref{alg:mcts}, there is only a single root node in the search tree at the beginning, and $t$ iterations of tree search are performed. In each iteration, we first utilize the \texttt{TreePolicy()} to select a tree node for expansion by balancing the exploration and exploitation. We then evaluate the newly expanded tree node with the \texttt{DefaultPolicy()}. Specifically, for the \texttt{TreePolicy()}, we start from the root node and select successive child nodes until a node $l$ with unexpanded child nodes is reached. During the node selection, the UCB (Upper Confidence Bound)~\cite{kocsis2006bandit} value is used to balance the exploration (gather more information about less-visited nodes) and exploitation (choose the optimal node based on existing information):
\begin{equation}
    \operatorname{Q}(n) + c \sqrt{\frac{2\ln{\operatorname{N}(n^{\prime})}}{\operatorname{N}(n)}}
\end{equation}
where $n$ indicates the current node, $n^{\prime}$ is its parent node, $Q()$ is the value function, $N()$ indicates the number of visit times, and $c$ is the exploration parameter. After reaching $l$, we then expand one child node for $l$ by randomly selecting an untried cutting plane $\mathcal{P}$ for the component $c^*$ and cutting $c^*$ with $\mathcal{P}$ (\cref{line:expand1,line:expand2,line:expand3}).

\subsection{Tree Node Evaluation} 

It's non-trivial to evaluate the expanded node, and one of the challenges is to compare decompositions with different number of cuttings (nodes at different depths). We thus employ a \texttt{DefaultPolicy()} to complete one playout, which leads to a fixed number of $d+1$ components with one-step greedy cuttings. Specifically, the greedy strategy first tries to cut the component $c^*$ with the largest concavity from the middle along three axis-aligned directions (\cref{line:default_cut}). By comparing the one-step concavity reduction (Equation~\ref{equ:concavity_deduction}), the strategy then keeps the best cutting results from the three trials. We repeatedly apply this greedy cutting strategy until we get $d+1$ decomposed components, and then calculate a score for the node. Please note that, in the tree search, we do not wait until all components are almost convex. Because then it could be much more time-consuming, and most cuttings may be achieved by the default policy, making the results less relevant to the searched tree nodes.

After applying the \texttt{TreePolicy()} and \texttt{DefaultPolicy()}, we get $d$ cutting planes and $d+1$ resulting components, where the first $l$ planes $\{\mathcal{P}_1, \cdots, \mathcal{P}_l\}$ are associated with a tree path starting from the root node to the leaf node, and the remaining planes $\{\mathcal{P}_{l+1}, \cdots, \mathcal{P}_{d}\}$ come from $\texttt{DefaultPolicy()}$. To evaluate the decomposition (\cref{line:score}), we not only assess the $d+1$ decomposed components after $d$ cuttings, but also examine the intermediate results. In this way, we can differentiate paths leading to similar final results and pick the path that achieves good results at a earlier stage. Specifically, the score for the set of cutting planes is calculated as:
\begin{equation}
 \operatorname{Quality}(\{\mathcal{P}_{1}, \cdots, \mathcal{P}_{d}\}) = \frac{1}{d} \sum_{i = 1}^{d} -\max_{j = 1}^{i+1} \operatorname{Concavity}(c_{ij})
 \label{equ:quality}
\end{equation}
where $c_{ij}$ represents one of the $i+1$ components after the $i$th cutting. At the end of each iteration, we update the scores for all nodes along the path (\cref{line:update}).

\subsection{Plane Refinement and Component Merging}
\label{sec:refinement}

\begin{wrapfigure}{l}{0.17\textwidth}
    \centering
    \includegraphics[width=0.2\textwidth]{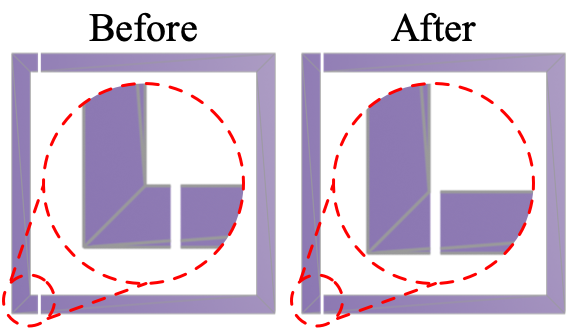}
\end{wrapfigure}
After completing an MCTS search ($t$ iterations), we take the optimal cutting plane of the root node (from the child with the highest score). Since we sample discretized equally-spaced planes as candidates, it's likely that the searched plane may not be the optimal one from a continuous space. As shown in the left inset, we thus locally refine the searched plane. Specifically, we find the $d$ cutting planes $\{\mathcal{P}_{1}, \cdots, \mathcal{P}_{d}\}$ that correspond to the optimal path in the search tree. We finetune the first plane $\mathcal{P}_{1}$ within a small range using a greedy ternary search, while other $d-1$ cutting planes and the score function (Equation~\ref{equ:quality}) remain the same. In this way, we can find high-resolution cutting planes without increasing the complexity of tree search. The refined plane is then used to cut $C$ into two parts. Please note that, in order to accelerate the tree search, we use $\operatorname{R_v}$ as the concavity within the MCTS. However, outside MCTS, we still use $\max(\operatorname{H_b}(\mathcal{S}), k\operatorname{R_v}(\mathcal{S}))$ to determine whether a component satisfies the concavity constraint and whether further cuttings are needed.

Since we recursively split a component into two, it's possible that among the set of decomposed components, there exist some components that could be merged to form a larger component that is still almost convex. We thus perform a post-processing to merge the generated components and further reduce the number of components. Specifically, we traverse all pairs of adjacent components and check the $\widetilde{\operatorname{Concavity}}$ of the merged component. If it's within the threshold $\epsilon$, we will replace the two components with the merged one. We repeat the process until no more components to merge.

\section{Evaluations}

\begin{figure*}[t]
\begin{center}
   \includegraphics[width=0.97\linewidth]{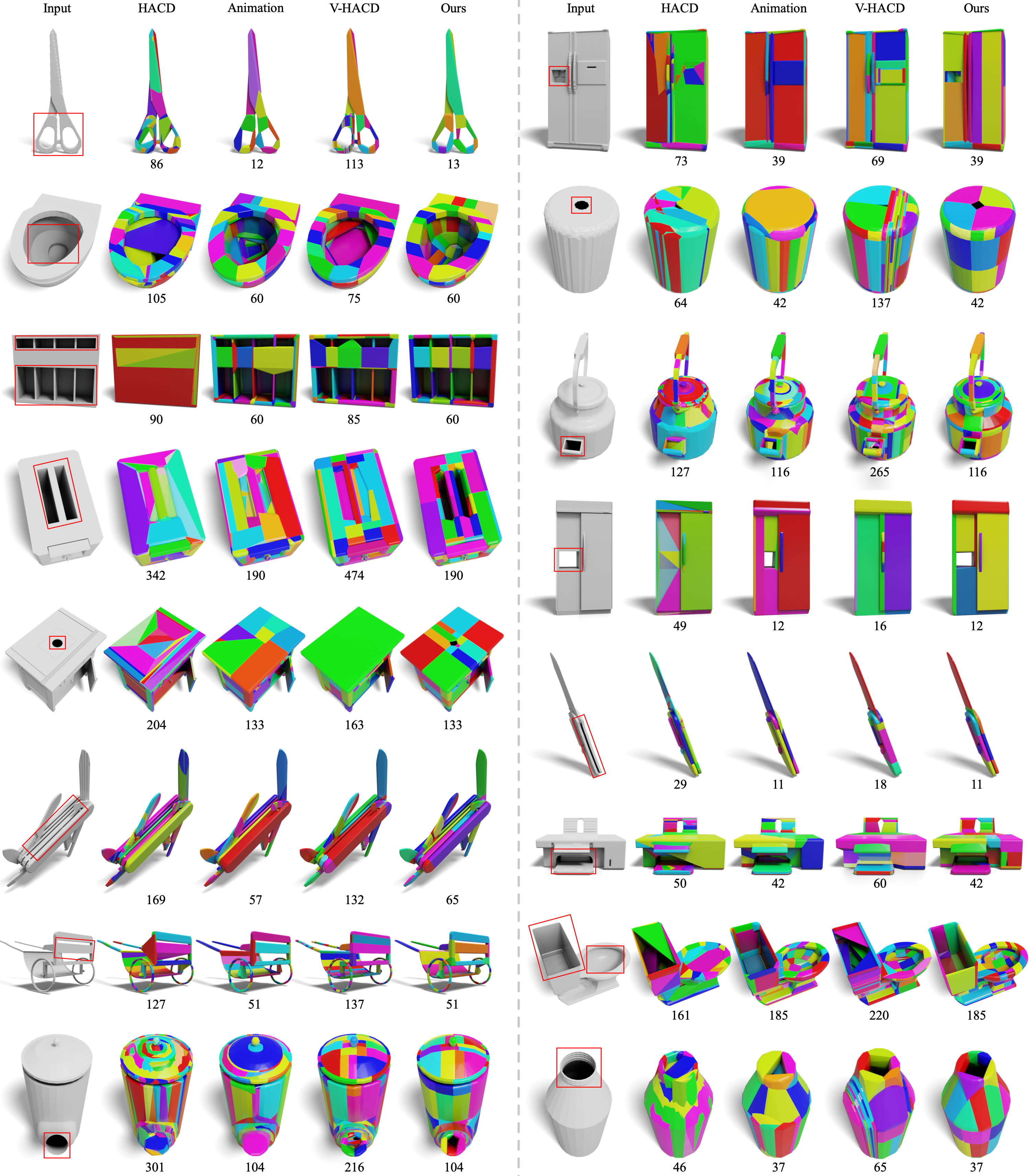}
\end{center}
  \caption{We compare our method with HACD, Animation, and V-HACD. Please zoom in for the details. The input shapes come from the PartNet-Mobility dataset. Each shape may consist of multiple parts (e.g., two blades of a scissor), and each part is decomposed individually. The red rectangles on the input shapes highlight the error-prone regions. The numbers below the results indicate the number of decomposed components. For both HACD and V-HACD, we set parameters to encourage fine-grained decomposition. For our method, we use a concavity threshold of 0.05. For Animation, we let the number of decomposed components equal our results. }
  \vspace{-4em}
\label{fig:exp-comparison}
\end{figure*}

\subsection{Comparing with Existing Methods}
We evaluate the methods on the V-HACD dataset~\cite{mamou2016volumetric} and PartNet-Mobility dataset~\cite{xiang2020sapien}. V-HACD dataset contains 61 shapes and most shapes are animals or humans. PartNet-Mobility dataset contains 2,346 shapes covering a wide range of indoor articulated objects (e.g., cabinets and scissors), which can be used for robotics simulation. Compared to the V-HACD dataset, shapes in the PartNet-Mobility dataset contain more complex inner structures and delicate details. It also requires higher quality decomposition to enable fine-grained object interaction. Each shape in the PartNet-Mobility dataset may contain multiple parts and we decompose each part individually.

\setlength{\tabcolsep}{3pt}
\begin{table}[t]
\small
  \centering
  \caption{Quantitative comparison on the V-HACD dataset and PartNet-Mobility dataset. For both HACD and V-HACD, we aim to compare the number of decomposed components. For Animation, we aim to match the number of decomposed components and compare the concavity scores. The runtime is in seconds. \\ $^1$ Animation is run with a different system configuration.}
    \begin{tabular}{c|c|ccc}
    \toprule
    dataset & method & \textbf{\# component} $\downarrow$ & concavity $\downarrow$ & runtime $\downarrow$ \\
    \midrule
    \multirow{2}[2]{*}{V-HACD} & HACD  &   57.6    &    0.118     & 67.2 \\
          & Ours  & \textbf{29.6}  &   0.084 & 201.0 \\
    \hline
    \multirow{2}[2]{*}{PartNetM} & HACD  &  33.5  &  0.414  & 268.9 \\
          & Ours  & \textbf{7.3}  &  0.204  & 194.4 \\
    \bottomrule
    \multicolumn{1}{c}{} & \multicolumn{1}{c}{} &       &             &  \\
    \toprule
    dataset & method & \textbf{\# component} $\downarrow$ & concavity $\downarrow$ & runtime $\downarrow$ \\
    \midrule
    \multirow{2}[2]{*}{V-HACD} & V-HACD & 60.2  &   0.067     & 192.1 \\
          & Ours  & \textbf{29.8}  & 0.044  & 201.9 \\
    \hline
    \multirow{2}[2]{*}{PartNetM} & V-HACD &   44.6  & 0.055 & 206.0 \\
          & Ours  & \textbf{20.1}  & 0.052  & 253.4 \\
    \bottomrule
    \multicolumn{1}{c}{} & \multicolumn{1}{c}{} &       &             &  \\
    \toprule
    dataset & method & \# component $\downarrow$ & \textbf{concavity} $\downarrow$ & runtime $\downarrow$ \\
    \midrule
    \multirow{2}[2]{*}{V-HACD} & Animation &  34.4   &  0.069 & 28.5$^1$ \\
          & Ours  & 34.5  & \textbf{0.049}  & 229.8 \\
    \bottomrule
    \end{tabular}
  \label{tab:comparison}
\end{table}

We compare our proposed method with existing approximate convex decomposition methods, HACD~\cite{mamou2009simple}, V-HACD~\cite{mamou2016volumetric}, and Animation~\cite{thul2018approximate}. It's non-trivial to compare different decomposition methods, since some methods use the concavity threshold as the termination rule, while other methods take the expected number of components as input. Moreover, different methods may have different concavity definitions. As a result, we compare our proposed method with each of the baseline method separately. Specifically, for comparison with HACD and V-HACD, we first run their methods by setting hyper-parameters that encourage as fine-grained decomposition as possible. After that, we calculate a concavity score for each of their generated decomposition solution:
\begin{equation}
    \operatorname{Score}(\mathcal{S}, \{\mathcal{CH}_1, \cdots, \mathcal{CH}_n\}) = \max_{i} \widetilde{\operatorname{Concavity}}(\mathcal{S} \cap \mathcal{CH}_i)
    \label{equ:exp_score}
\end{equation}
where $\{\mathcal{CH}_1, \cdots, \mathcal{CH}_n\}$ indicates the set of generated convex hulls for the input shape $\mathcal{S}$, and $\mathcal{S} \cap \mathcal{CH}_i$ calculates the intersection between the input solid shape and the $i$th convex hull. We then utilize that score as the concavity threshold for running our method. In this way, our method will generate decomposition with finer details, and we aim to compare the number of decomposed components. 

For comparison with Animation, we first run our method with a fixed concavity threshold of 0.05. We then run Animation for each shape to generate the same number of components as our results. After that, we can fairly compare the concavity scores (Equation~\ref{equ:exp_score}) of the two methods. Since Animation didn't release their code due to commercial reasons, we ask the authors to help us run the method on their machine. We quantitatively compare our method and Animation only on the V-HACD dataset, since PartNet-Mobility dataset is too large.

As shown in Table~\ref{tab:comparison}, our method outperforms HACD and V-HACD in terms of the number of components on both datasets with a large margin. At the same time, the mean concavity scores of our results are even lower than that of the two methods. As for comparison with Animation, both methods generate almost the same number of components for each shape. However, we achieve lower concavity scores which indicate the finer decomposition results by our method. We run methods with a single CPU thread (except for Animation) and we share similar runtime with HACD and V-HACD. Animation is more time-efficient, since its implementation has been carefully optimized with industrial codes.

Figure~\ref{fig:exp-comparison} shows the qualitative comparison on the PartNet-Mobility dataset, where many shapes need fine-grained decomposition to enable downstream object interaction. For example, the inside ring of the scissors should be large enough, so that an agent can grab them, the slots of the toaster and the spouts of the kettles should not be filled so that they can work properly. However, decomposition results from existing methods may fail to preserve the original shape's functionality. Among the three baselines, HACD produces the worst results since it only considers the difference between the boundary surfaces while ignoring the interior structures. It fills the inner space for most shapes. V-HACD and Animation produce better results but still suffer from the hole filling issues to some extent. Both methods utilize the volume difference as the concavity metric and may ignore fine-grained structures or introduce some thin-planar components to fill the holes, since those errors do not receive a large penalty from the volume-based concavity. Instead, by leveraging our collision-aware concavity, our method keeps most structures of the input shapes. Also, the numbers of our decomposed components are smaller than those of HACD and V-HACD, which may speed up the downstream applications. 

\subsection{Ablation studies}

\begin{figure}[t]
\begin{center}
   \includegraphics[width=\linewidth]{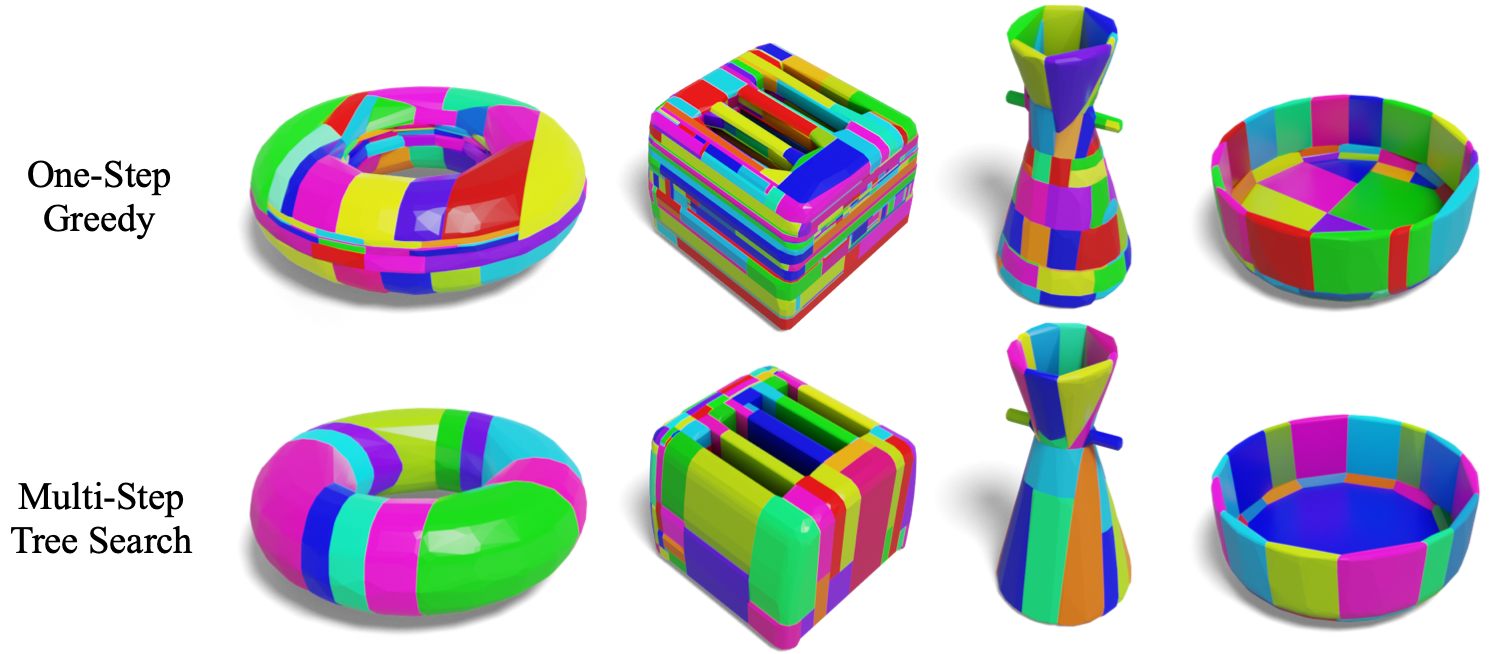}
\end{center}
  \caption{First row: results from a one-step greedy strategy with our proposed concavity metric. Second row: results from our method with multi-step tree search. Both methods are tested with the same concavity threshold.}
\label{fig:mcts_greedy_exp}
\end{figure}

\begin{figure*}
     \centering
     \begin{subfigure}[b]{0.24\textwidth}
         \centering
         \includegraphics[width=\textwidth]{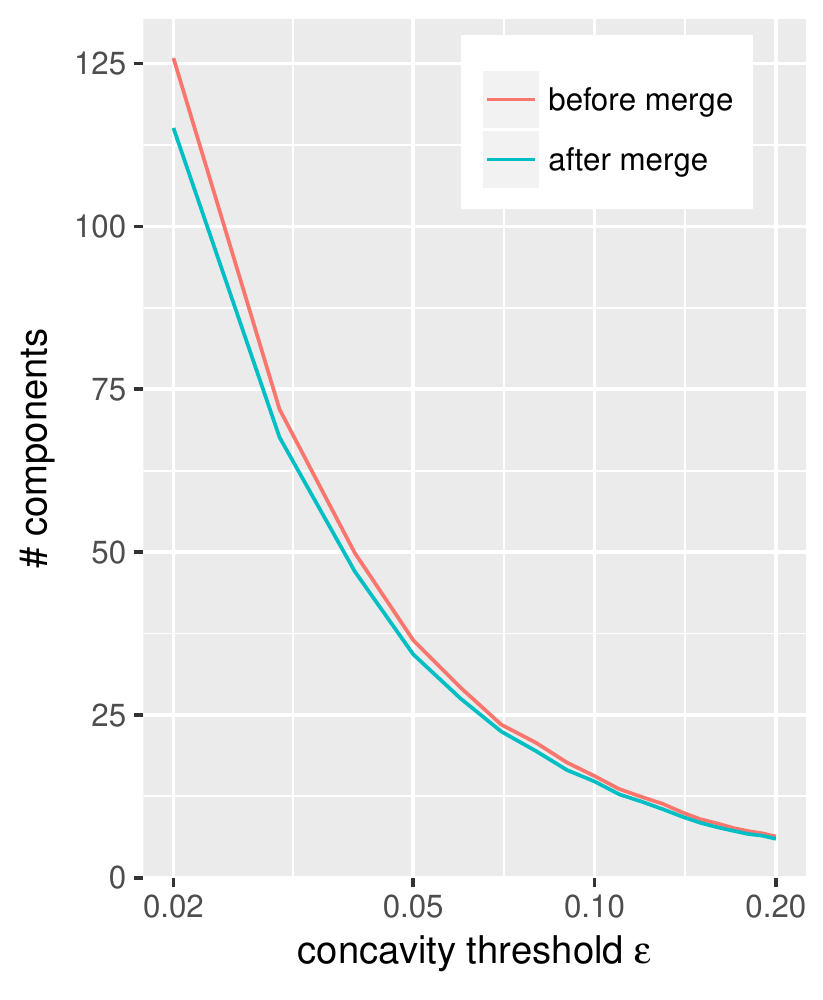}
         \caption{}
         \label{fig:ablation_threshold}
     \end{subfigure}
     \hfill
     \begin{subfigure}[b]{0.24\textwidth}
         \centering
         \includegraphics[width=\textwidth]{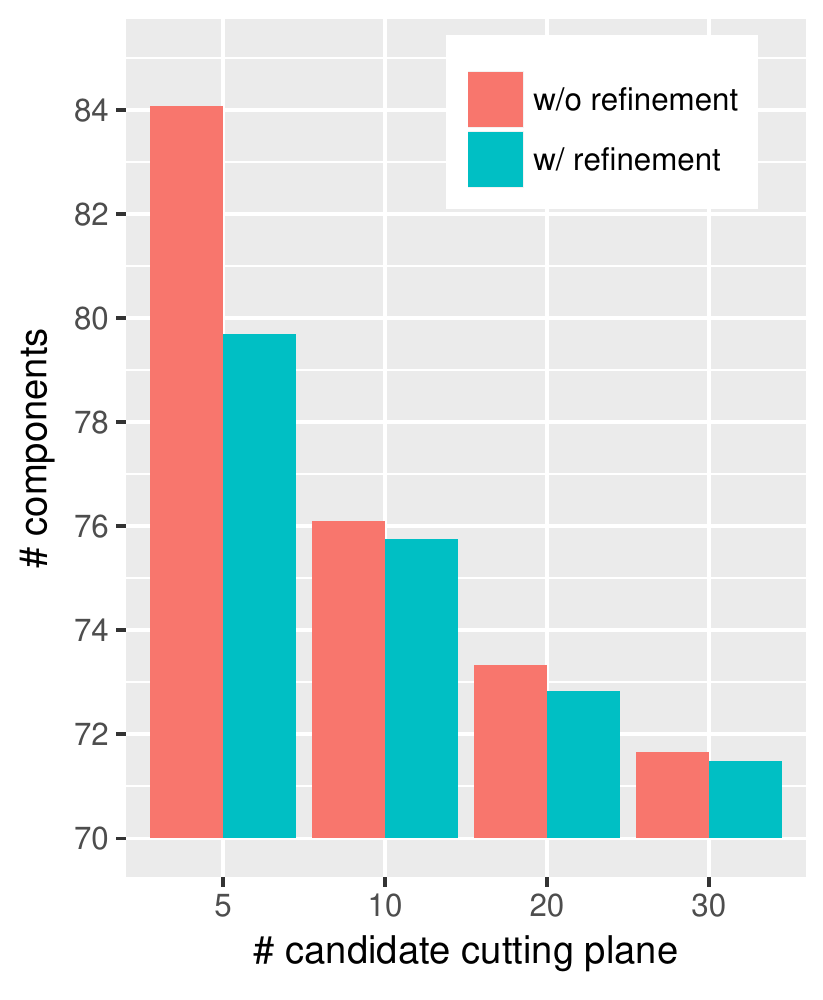}
         \caption{}
         \label{fig:ablation_node}
     \end{subfigure}
     \hfill
     \begin{subfigure}[b]{0.24\textwidth}
         \centering
         \includegraphics[width=\textwidth]{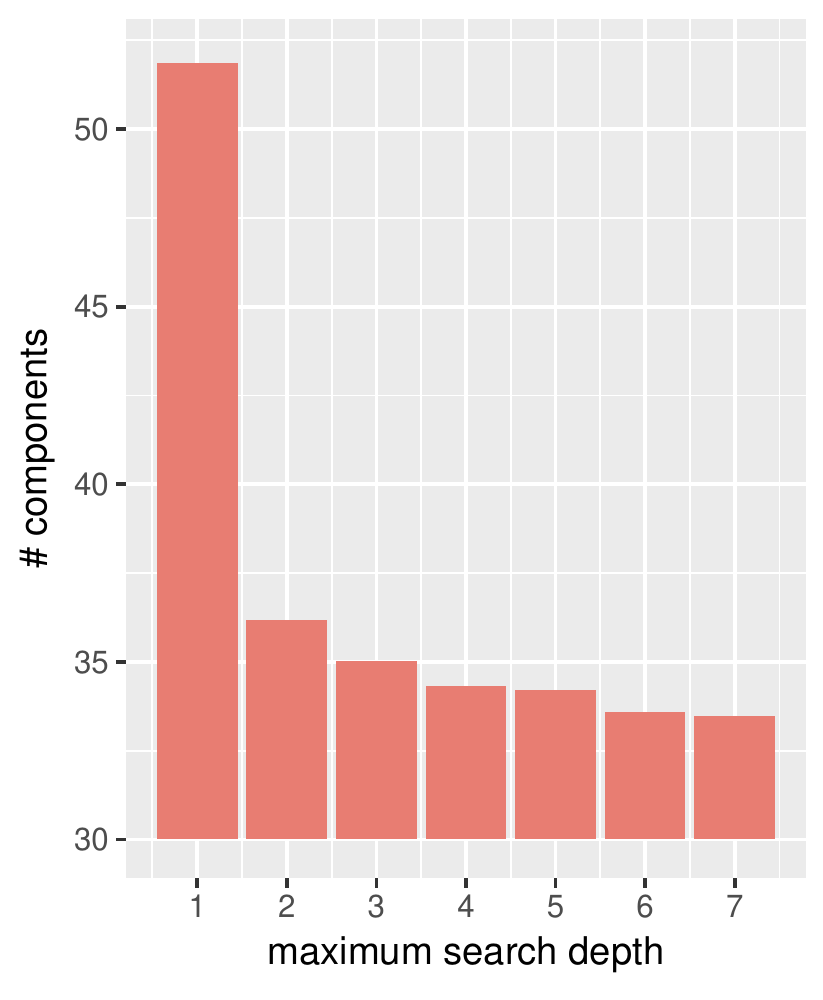}
         \caption{}
         \label{fig:ablation_depth}
     \end{subfigure}
         \hfill
     \begin{subfigure}[b]{0.24\textwidth}
         \centering
         \includegraphics[width=\textwidth]{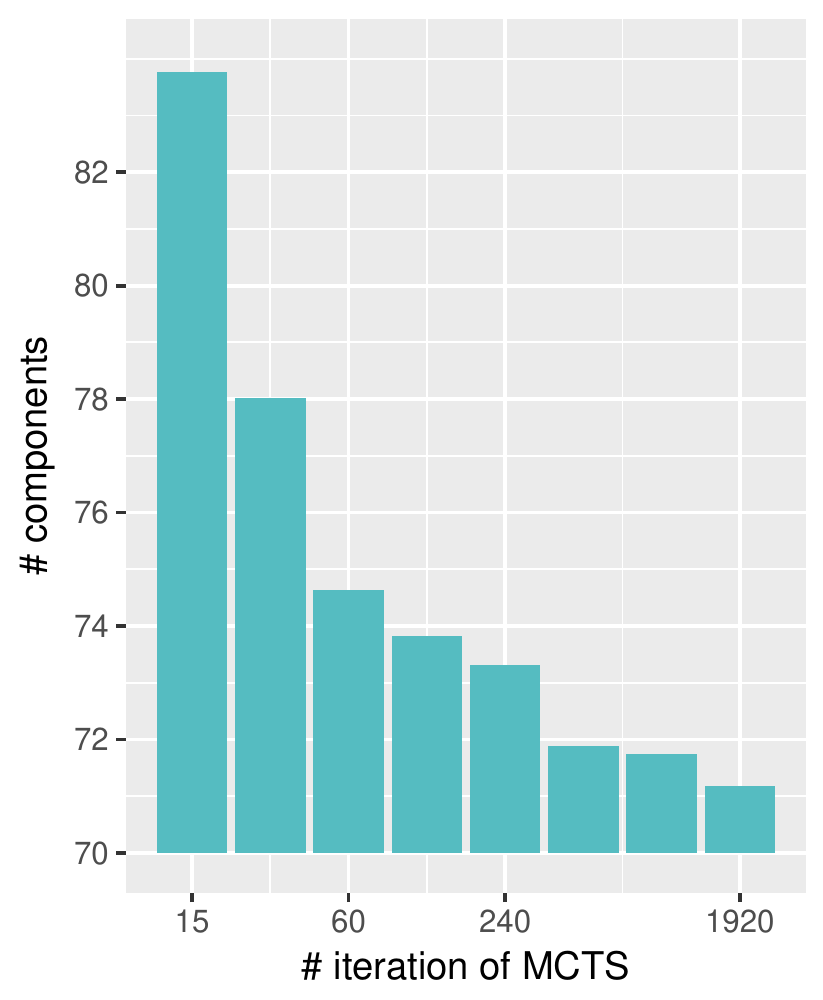}
         \caption{}
         \label{fig:ablation_iteration}
     \end{subfigure}
        \caption{Ablation studies: (a) the concavity threshold $\epsilon$ and the post-processing merge, (b) the number of sampled cutting plane candidates from each axis-aligned direction (i.e.,  $m$), (c) the maximum search depth $d$ in each MCTS, (d) the number of iterations in each MCTS (i.e., $t$). For all figures, the y-axis represents the number of decomposed component under each setting.}
        \label{fig:ablation}
\end{figure*}

\begin{figure*}[t]
\begin{center}
   \includegraphics[width=\linewidth]{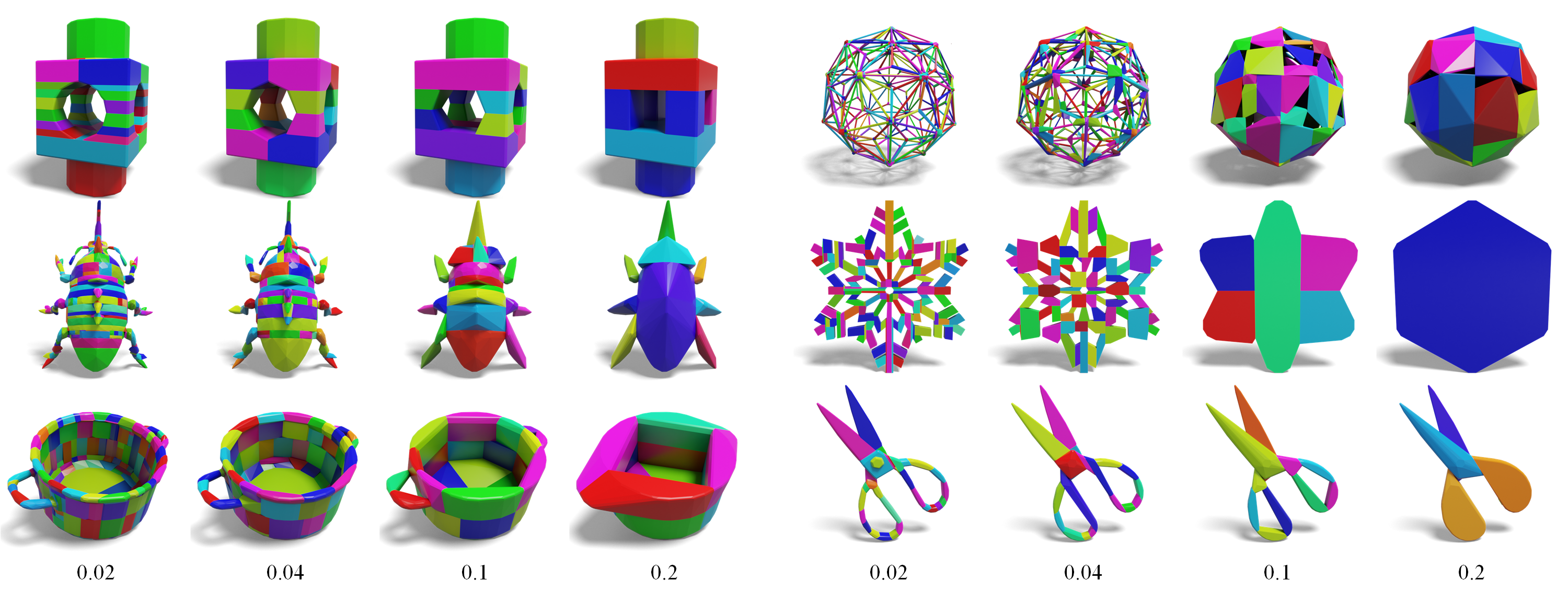}
\end{center}
  \caption{Comparison of different concavity thresholds. For each example, we show the decomposition results under different concavity thresholds ranging from 0.02 to 0.2. Users can intuitively balance the level of detail and the number of components by adjusting the concavity threshold $\epsilon$.}
\label{fig:threshold}
\end{figure*}

\paragraph{\textbf{One-step greedy vs. multi-step tree search}} 

\setlength{\tabcolsep}{5pt}
\begin{table}[t]
  \centering
    \caption{Quantitative comparison between a one-step greedy baseline and our method with multi-step tree search on the V-HACD dataset.}
    \begin{tabular}{cc|cc}
    \toprule
    \multicolumn{2}{c|}{One-Step Greedy} & \multicolumn{2}{c}{Multi-Step Tree Search} \\
    \hline
    \multicolumn{1}{l}{\# component $\downarrow$ } & \multicolumn{1}{l|}{runtime $\downarrow$} & \multicolumn{1}{l}{\# component $\downarrow$} & \multicolumn{1}{l}{runtime $\downarrow$} \\
    \midrule
      49.9   & 271.7  & \textbf{34.5}  & \textbf{229.8} \\
    \bottomrule
    \end{tabular}
  \label{tab:greedy_mcts}
\end{table}

To examine the benefit introduced by the multi-step tree search, we construct a counterpart greedy baseline which utilizes our proposed concavity metric and directly search for the best one-step cutting plane that leads to the minimum resulting concavity (Equation~\ref{equ:concavity_deduction}). We compare the one-step greedy baseline and our proposed method with the same concavity threshold 0.05. The quantitative results on the V-HACD dataset are shown in Table~\ref{tab:greedy_mcts}, where the one-step greedy baseline generates much more components than our multi-step tree search version. Moreover, since the multi-step tree search version reduces the number of rounds (fewer parts) and utilizes a simplified concavity calculation in the tree search, it is even faster than the greedy baseline. 

As shown in Figure~\ref{fig:mcts_greedy_exp}, the one-step greedy algorithm may be short-sighted and generate more components, while the results with multi-step tree search are more reasonable. For example, when searching for the first cutting plane of the torus (first from left), either vertically or horizontally cutting can lead to sub-parts with the same concavity score, and the greedy algorithm will thus randomly select the first cutting plane. However, horizontal cuttings will lead to more components in the final results. Similarly, for the bottle cap example (first from right), the one-step greedy algorithm will not cut off the bottom in the first step because it will even increase the concavity score. However, by leveraging the multi-step tree search, we can find that cutting off the bottom in the first step can avoid the bottom being divided into unnecessary parts.

\paragraph{\textbf{Impact of the concavity threshold $\epsilon$.}} 
Our method terminates when the concavities of all decomposed components are less than a pre-defined threshold $\epsilon$. The concavity threshold $\epsilon$ thus balances the level of details and the number of decomposed components. As shown in Figure~\ref{fig:ablation_threshold} and Figure~\ref{fig:threshold}, when we decrease the concavity threshold $\epsilon$, the algorithm generates more components to preserve the details and the generated convex hulls are much closer to the original shape. When we increase the concavity threshold $\epsilon$, we generate fewer components to approximate the global structure of the original shape and may lose some of the details. The variation is more significant when the concavity is relatively small. 

We also want to point out that compared to the volume-based concavity, our proposed concavity metric measures the distance. One can interpret the threshold as the degree to which the original shape becomes thicker, which may be more intuitive for users to adjust the threshold and achieve their desired decomposition. In contrast, the volume-based concavity may not correspond to such an intuitive interpretation, and the change caused by adjusting the threshold may be less predictable.

\paragraph{\textbf{Impact of hyper-parameters in the tree search.}}

We study the impact of the hyper-parameters in the multi-step tree search by fixing other hyper-parameters and the concavity threshold. The ablation results are shown in the Figure~\ref{fig:ablation}. (i) We sample $m$ candidate cutting planes from each axis-aligned direction. By sampling more candidate planes, we achieve more precise cuttings, which are much closer to the optimal location. As shown in Figure~\ref{fig:ablation_node}, a larger $m$ thus leads to fewer components. (ii) We limit the maximum depth of the search tree to $d$ and evaluate each tree node by generating $d+1$ components. A larger $d$ enables the algorithm to analyze cuttings in further steps and achieve a more precise tree node evaluation. As shown in Figure~\ref{fig:ablation_depth}, a larger $d$ leads to a better performance generally. Moreover, we find that seeing one step further (i.e., $d = 2$) introduces the most significant gain. (iii) As shown in Figure~\ref{fig:ablation_iteration}, searching for more iterations leads to better solutions, since a larger number of iterations $t$ means expanding more nodes, exploring more cutting combinations, and more accurate evaluation for the tree nodes. However, increasing the three hyper-parameters causes a longer search time. As a result, there are trade-offs between the decomposition quality and the runtime.

\paragraph{\textbf{Refinement, Merging, and Cutting Directions.}}
The cutting plane refinement aims to find a better position in the continuous local neighborhood of the searched discretized candidates, thus enabling a more precise cutting. As shown in Figure~\ref{fig:ablation_node}, when the number of candidate cutting planes increases, the improvement brought by the refinement becomes smaller due to the narrower gap between two adjacent candidates.

\begin{figure}[t]
\begin{center}
   \includegraphics[width=\linewidth]{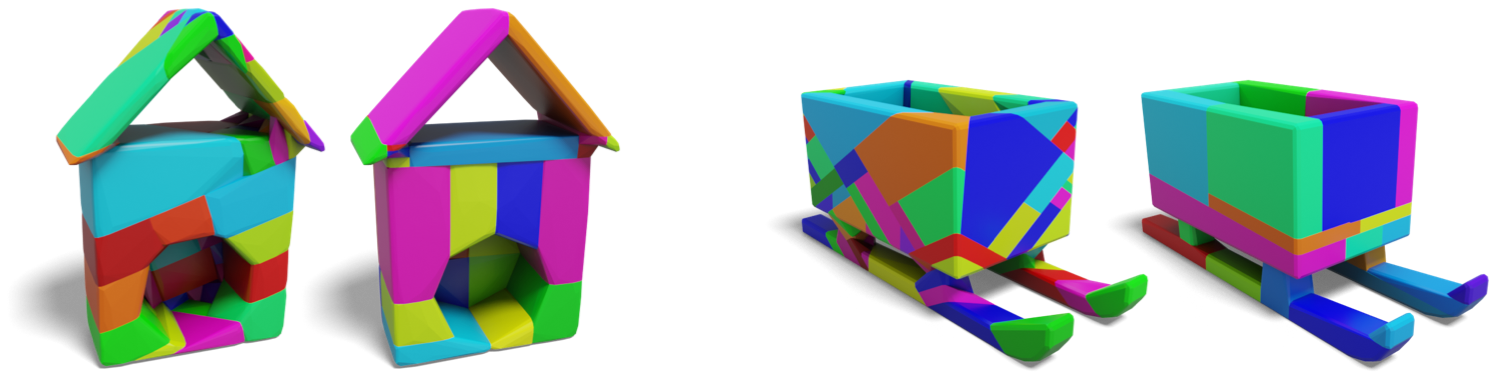}
\end{center}
  \caption{Comparison of different cutting directions. In each pair, the left one indicates cutting with a set of random axes, while the right one indicates cutting with the principal axes computed by PCA.}
\label{fig:pca}
\end{figure}

After decomposition, we merge components as post-processing to further reduce the number of components. As shown in Figure~\ref{fig:ablation_threshold}, when the concavity threshold is smaller, the shape is decomposed into more pieces, and the component merging can reduce more redundant divisions.

Since we sample cutting planes from three mutually orthogonal directions as V-HACD, the selection of the cutting directions may have a great influence on the final results in some cases, as shown in Figure~\ref{fig:pca}. By specifying a set of good axes or computing principal axes by PCA, we may generate fewer components.

\subsection{Application}

\begin{figure}[t]
\begin{center}
   \includegraphics[width=\linewidth]{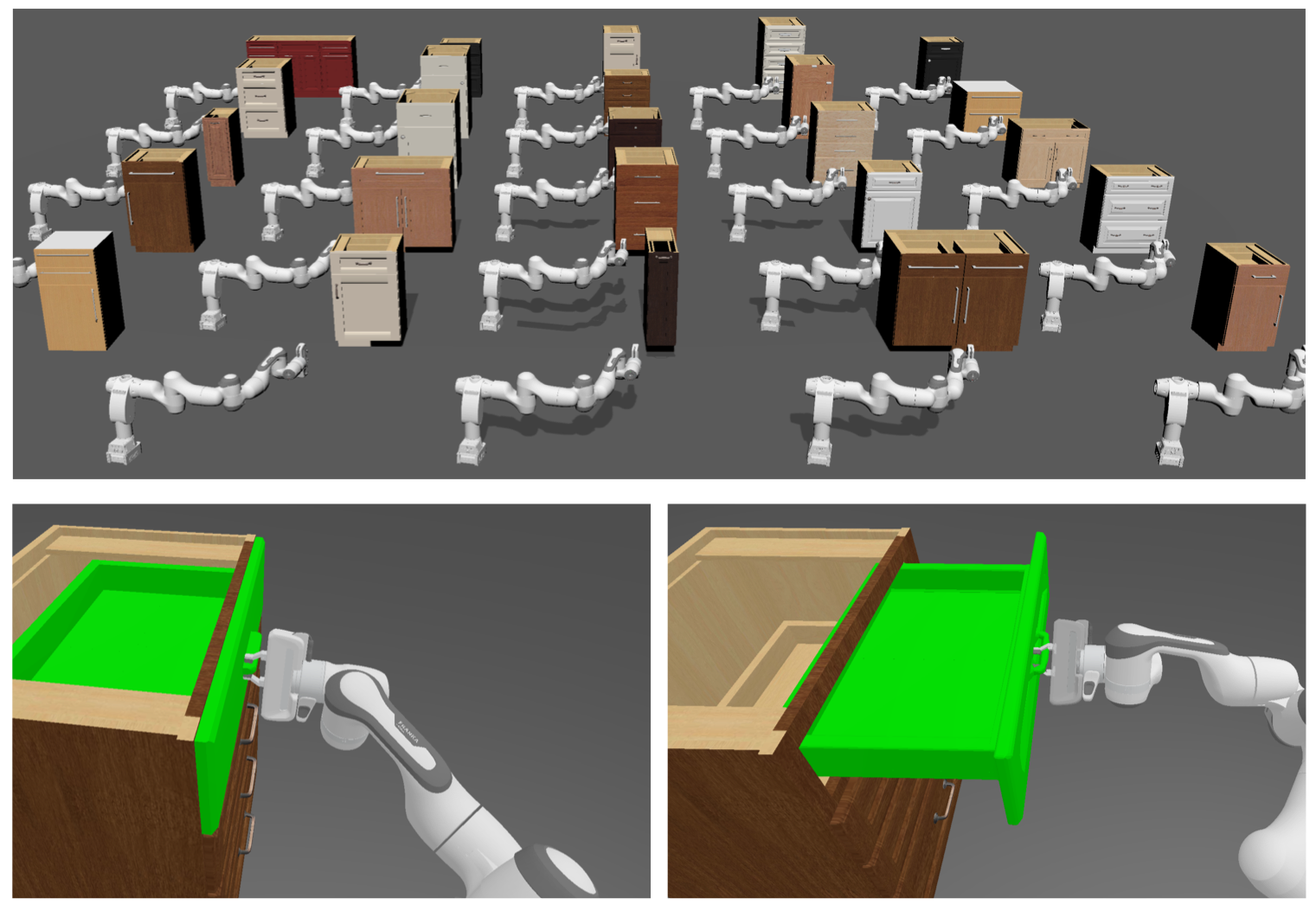}
\end{center}
  \caption{Top: We train RL agents to open 49 drawers of 25 cabinets in a physics simulator. Bottom left: Using decomposition results of V-HACD as collision shapes. The collision shape of the drawer is highlighted in green, and the hole of the handle is filled (zoom in for details). Bottom right: Using our decomposition results as collision shapes. We preserve fine-grained details of the handle.}
\label{fig:open_cabinet}
\end{figure}

An important application of convex decomposition is to provide collision shapes for physics simulators that perform extensive collision detection. On the one hand, we aim to approximate the shape with a small number of convex components, thereby speeding up the collision detection. On the other hand, we want the decomposed components to closely match the original shape, so that the functionality of the object is not compromised.

In this experiment, we compare two sets of collision shapes generated by our method and V-HACD. Specifically, we load 25 cabinets into SAPIEN~\cite{xiang2020sapien}, a physics simulator. We utilize our method and V-HACD to generate a collision shape (i.e., an assembly of convex components) for each part (e.g., a drawer or a body), respectively. As shown in Figure~\ref{fig:open_cabinet}, the collision shapes by our method preserve fine-grained details of the handles, while the collision shapes by V-HACD fill the holes of the handles even a tiny threshold is used.

\begin{table}[t]
\small
  \centering
  \caption{Results of the OpenCabinetDrawer task. We compare using different decomposition results as the collision shapes.}
    \begin{tabular}{c|cc}
    \toprule
          & V-HACD & Ours \\
    \midrule
    Successfully Opened Drawer & 49\%  & \textbf{80\%} \\
    \bottomrule
    \end{tabular}
  \label{tab:open_drawer}
\end{table}

We train SAC~\cite{haarnoja2018soft} (a reinforcement learning algorithm) agents to control a robot arm to open the drawers. Specifically, there are 49 drawers from the 25 cabinets. We train an individual SAC agent from scratch for each drawer with $10^6$ time steps per trial. Please refer to~\cite{mu2021maniskill} for other training details. Since reinforcement learning algorithms are not guaranteed to converge to the optimum every run, if we open a drawer in 5 trials, we regard it as a success case. We report the result in Table~\ref{tab:open_drawer}. 

By using more accurate collision shapes generated by our method, the RL agents achieve a much higher success rate. We observe that, when using our collision shapes, which preserve the fine-grained details (e.g., holes) of the handles, the robot arm is easier to form a shape-closure grasp, which is more robust. However, when using V-HACD's collision shapes, the robot arm easily slips off the handles, since they fill the holes.

\section{Discussion}

We propose a novel approximate convex decomposition method that differs from prior approaches in three folds: (a) we introduce a novel collision-aware concavity metric that better examines the shapes from both the boundary and the interior. It preserves fine-grained structures of the input shape and enables delicate object interaction in downstream applications. (b) we decompose the shape by efficiently cutting the meshes. It ensures intersection-free components and avoids discretization artifacts. (c) we utilize multi-step tree search to find globally better cutting planes, leading to fewer decomposed components. 
 
We currently adopt many simplifications due to the runtime consideration. In the future, we would like to optimize our implementation further (e.g., utilize parallelization). We may also employ deep neural networks to help evaluate a decomposition solution more efficiently and accurately. Moreover, we can explore a smarter way to pick the cutting directions and set adaptive thresholds for different parts to reduce the number of decomposed components.

\begin{acks} 
This work is supported in part by gifts from Adobe, Qualcomm, and Kingstar. We would like to thank Lubor Ladicky and Daniel Thul for running the Animation code, thank Abdulaziz Almuzairee, Xiaoshuai Zhang, Jiayuan Gu, Fanbo Xiang, and Songfang Han for proofreading the manuscript. 
\end{acks}

\bibliographystyle{ACM-Reference-Format}
\bibliography{references}

\appendix

\section{Proof for Theorem 1}
In the paper, we propose a surrogate term $\operatorname{R_v}(\mathcal{S})$ to accelerate the computation of $\operatorname{H_i}(\mathcal{S})$ and provide a theoretical guarantee:

\begin{theorem} For every solid shape $\mathcal{S}$, we have
$$\sqrt{2}\max(\operatorname{H_b}(\mathcal{S}), \operatorname{R_v}(\mathcal{S})) \geq \max(\operatorname{H_b}(\mathcal{S}), \operatorname{H_i}(\mathcal{S}))$$
\end{theorem}

Here we give the detailed proof for the theorem. Recall that the Hausdorff distance for two point sets $A$ and $B$ is calculated as:
\begin{equation}
    \operatorname{H}(A, B) = \max \{\sup_{a \in A}d(a, B), \sup_{b \in B}d(b, A)\}
\end{equation}

where $d(x, Y) = \inf_{y \in Y}d(x, y)$ and $d(x, y)$ indicates the Euclidean distance between the two points.

When calculating $\operatorname{H_i}(\mathcal{S})$, the two point sets are sampled from the interior of the solid shape $\mathcal{S}$ and its convex hull $\operatorname{CH}(\mathcal{S})$. We denote them as $P$ and $Q$, respectively:
\begin{equation}
    P = \operatorname{Sample}(\operatorname{Int} \mathcal{S})
\end{equation}
\begin{equation}
    Q = \operatorname{Sample}(\operatorname{Int} \operatorname{CH}(\mathcal{S}))
\end{equation}
In our proof, we assume that the interior doesn't exclude the boundary surface, which is slightly different from the usual definition. Also, we assume that $\operatorname{Sample}(T)$ cover all points in $T$ (infinite sampled points). 

Since $\mathcal{S}$ is contained by $\operatorname{CH}(\mathcal{S})$, $d(p, Q) = 0$ for all $p \in P$.  We can thus simplify $\operatorname{H_i}(\mathcal{S})$ as:
\begin{equation}
    \operatorname{H_i}(\mathcal{S}) = \sup_{q \in Q} d(q, P)
\end{equation}

We know that there exists a pair of points $p^* \in P$ and $q^* \in Q$, such that $\operatorname{H_i}(\mathcal{S}) = d(q^*, P) = d(p^*, q^*) $. We prove the theorem by enumerating all possible locations of $p^*$ and $q^*$, which can be divided into four cases.

\hfill 

\noindent\textbf{Case 1: $q^*$ lies inside of $\mathcal{S}$.}

In this case, $\operatorname{H_i}(\mathcal{S}) = d(q^*, P) = 0$, and the theorem holds.

\hfill 

\noindent\textbf{Case 2: $p^*$ is not on the boundary surface of $\mathcal{S}$.}

\begin{figure}[t]
\begin{center}
  \includegraphics[width=0.5\linewidth]{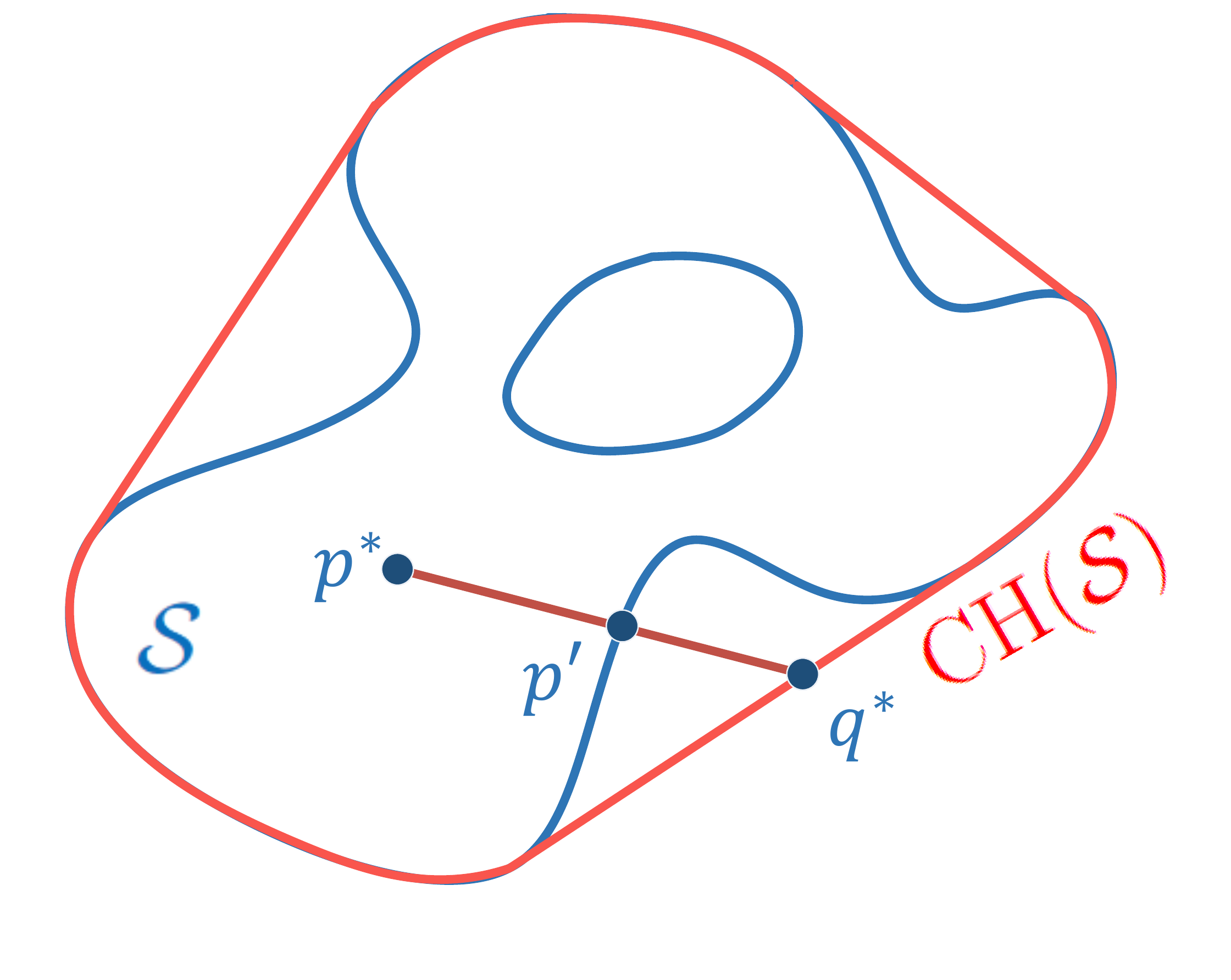}
\end{center}
  \caption{Counter example of Case 2. $p^*$ must be on the boundary surface of $\mathcal{S}$.}
\label{fig:point_inside}
\end{figure}

This case is impossible. Since $q^*$ lies outside of $\mathcal{S}$ (not Case 1), there must exist another point $p^{\prime}$ on the boundary surface of $\mathcal{S}$, such that $d(q^*, p^{\prime}) < d(q^*, p^*)$, which contradicts $d(q^*, P) = d(q^*, p^*)$. See Figure~\ref{fig:point_inside} for a illustration.

\hfill 

\noindent\textbf{Case 3: $p^*$ is on the boundary surface of $\mathcal{S}$ and $q^*$ is on the boundary surface of $\operatorname{CH}(\mathcal{S})$.}

In this case, we have $\operatorname{H_i}(\mathcal{S}) = \operatorname{H_b}(\mathcal{S})$, and the theorem holds.

\hfill 

\noindent\textbf{Case 4: $p^*$ is on the boundary surface of $\mathcal{S}$ and $q^*$ is not on the boundary surface of $\operatorname{CH}(\mathcal{S})$.}

\begin{figure}
     \centering
     \begin{subfigure}[b]{0.22\textwidth}
         \centering
         \includegraphics[width=\textwidth]{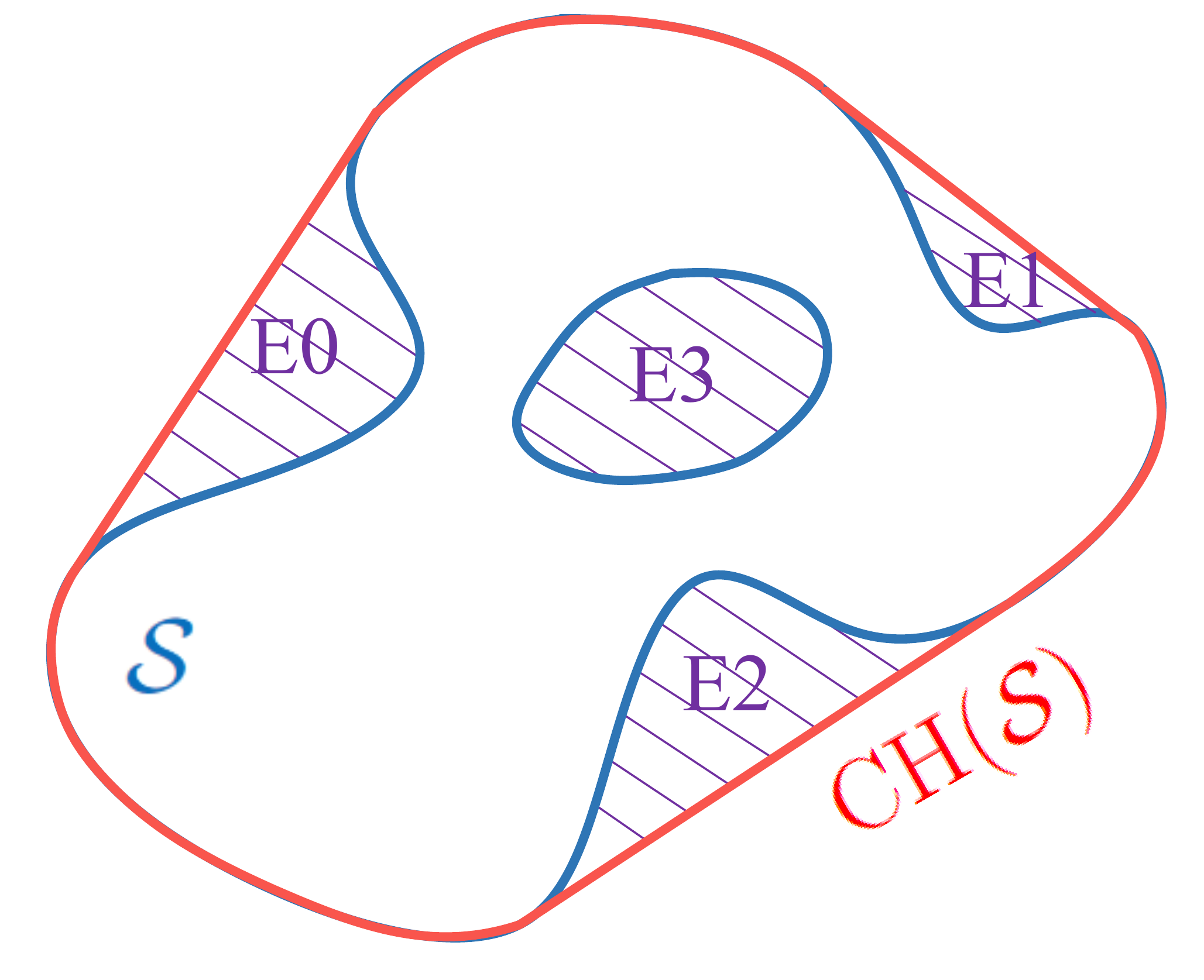}
         \caption{}
         \label{fig:intv_shape}
     \end{subfigure}
     \hfill
     \begin{subfigure}[b]{0.22\textwidth}
         \centering
         \includegraphics[width=\textwidth]{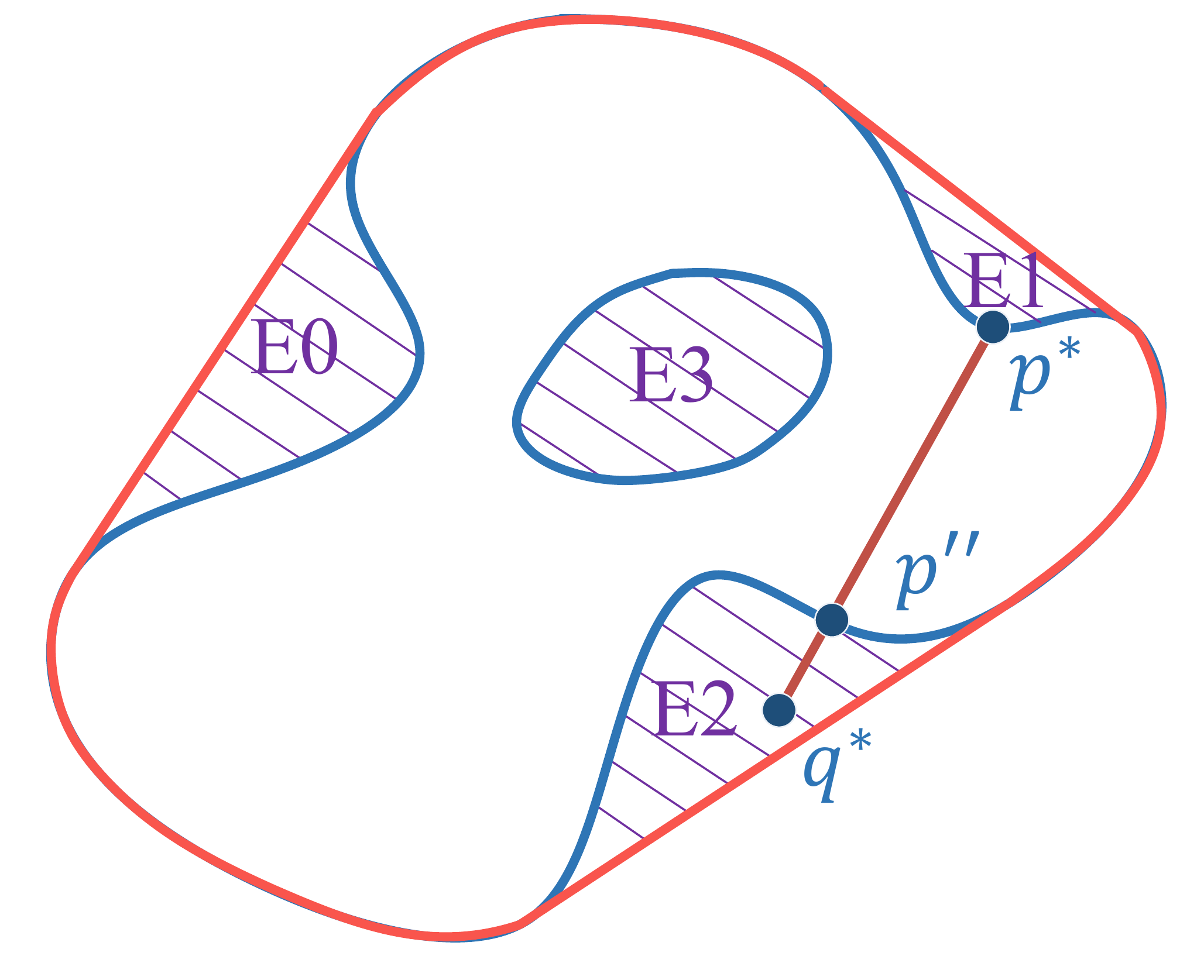}
         \caption{}
         \label{fig:same_block}
     \end{subfigure}
        \caption{Illustration of the interval space. Blue lines indicate the solid shape $\mathcal{S}$, and the red lines indicate its convex hull $\operatorname{CH}(\mathcal{S})$. (a) The shaded area shows the interval space, which consists of four connected regions $E_i$. (b) If $p^*$ and $q^*$ lie in different $E_i$, the segment connecting $p^*$ and $q^*$ will intersect with the boundary surface of $\mathcal{S}$ at another point $p^{\prime\prime}$, and $d(p^{\prime\prime}, q^*) < d(p^*, q^*)$.}
        \label{fig:interval_shapes}
\end{figure}

$q^*$ must lie within $\operatorname{CH}(\mathcal{S}) - \mathcal{S}$, the space outside of $\mathcal{S}$ but inside of $\operatorname{CH}(\mathcal{S})$. As shown in Figure~\ref{fig:intv_shape}, we call the space between $\mathcal{S}$ and $\operatorname{CH}(\mathcal{S})$ as the \textbf{interval space}, which may consists of multiple connected regions $E_i$. We denote the connected region containing both $p^*$, $q^*$ as $E_*$.

Note that $(p^*, q^*)$ cannot locate on different $E_i$. Otherwise, as shown in Figure~\ref{fig:same_block}, there must exist another point $p^{\prime\prime}$ on the boundary surface of $\mathcal{S}$, such that $d(q^*, p^{\prime\prime}) < d(q^*, p^*)$, which contradicts $d(q^*, P) = d(q^*, p^*)$.

Before we talk about the connection between $\operatorname{R_v}(\mathcal{S})$ and $\operatorname{H_i}(\mathcal{S})$, we first construct a \textbf{maximum inscribed sphere} within $E_{*}$ centered at $q^*$. We denote this sphere as $\Phi$ and its radius as $r$. We know that:
\begin{equation}
 \operatorname{Vol}(\operatorname{CH}(\mathcal{S})) - \operatorname{Vol}(\mathcal{S}) \geq  \operatorname{Vol}(E_*) \geq \operatorname{Vol}(\Phi)
\end{equation}

Combined with the definition of $\operatorname{R_v}(\mathcal{S})$, we can infer that $\operatorname{R_v}(\mathcal{S}) \geq r$.

\hfill

\noindent\textbf{Case 4.1: $\Phi$ does not intersect with the boundary surface of $\operatorname{CH}(\mathcal{S})$.}

In this case, $\Phi$ must intersect with the boundary surface of $\mathcal{S}$ at $p^*$. Otherwise, there exist another point $p^{\prime\prime\prime}$ on the boundary surface of $\mathcal{S}$, such that $d(q^*, p^{\prime\prime\prime}) < d(q^*, p^*)$, which contradicts $d(q^*, P) = d(q^*, p^*)$. As a result, $d(q^*, p^*)$ is equal to the radius $r$. Since  $\operatorname{R_v}(\mathcal{S}) \geq r$, we have:

\begin{equation}
       r = d(p^*, q^*) = \operatorname{H_i}(\mathcal{S})  \leq \operatorname{R_v}(\mathcal{S})
\end{equation}
The theorem holds.

\hfill

\noindent\textbf{Case 4.2: $\Phi$ intersects with the boundary surface of $\operatorname{CH}(\mathcal{S})$.}

\begin{figure}
     \centering
     \begin{subfigure}[b]{0.25\textwidth}
         \centering
         \includegraphics[width=\textwidth]{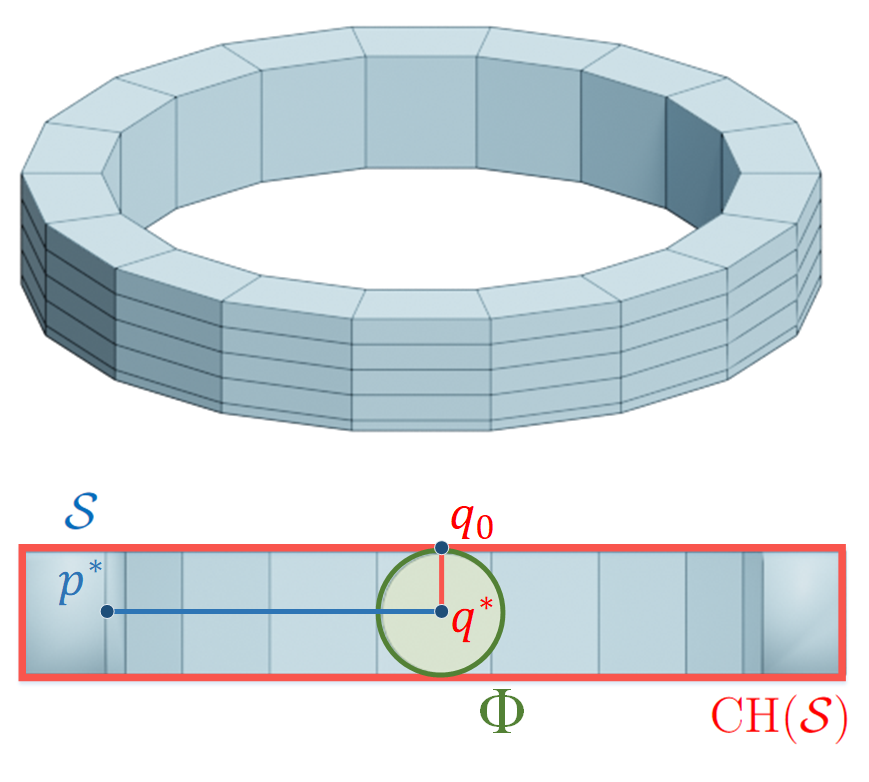}
         \caption{}
         \label{fig:hi_r}
     \end{subfigure}
     \hfill
     \begin{subfigure}[b]{0.22\textwidth}
         \centering
         \includegraphics[width=\textwidth]{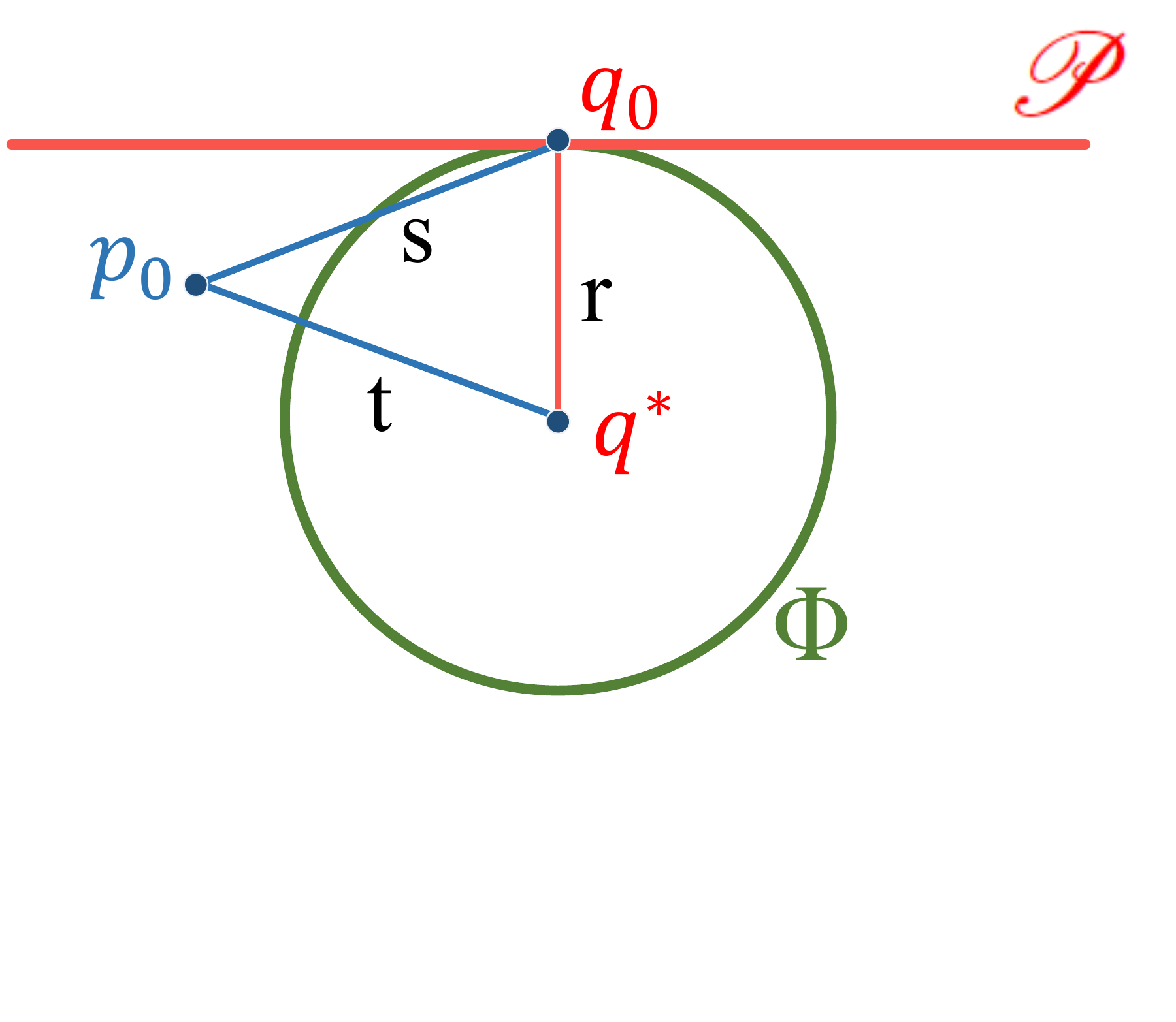}
         \caption{}
         \label{fig:distance}
     \end{subfigure}
        \caption{ (a) An example that the maximum inscribed sphere $\Phi$ is bounded by the boundary surface of $\operatorname{CH}(\mathcal{S})$ and $\operatorname{H_i}(\mathcal{S})$ does not equal to the radius $r$. (b) Illustration of the proof in Case 4.2.}
        \label{fig:dist}
\end{figure}

In this case, $\operatorname{H_i}(\mathcal{S})$ does not equal to the radius $r$, since the maximum inscribed sphere $\Phi$ is bounded by the boundary surface of $\operatorname{CH}(\mathcal{S})$ and it may fail to touch any point on the boundary surface of $\mathcal{S}$. Figure~\ref{fig:hi_r} shows such an example.

As shown in Figure~\ref{fig:distance}, we denote one of the intersection point as $q_0$. We know that $q_0$ is on the boundary surface of $\operatorname{CH}(\mathcal{S})$ and we have $d(q^*, q_0) = r$. We also find $q_0$'s nearest point on the boundary surface of $\mathcal{S}$ and denote it as $p_0$. We denote $d(p_0, q_0)$ as $s$ in the figure, and we know that $s \leq \operatorname{H_b}(\mathcal{S})$. We want to calculate the distance between $p_0$ and $q^*$, which is denoted as $t$ in the figure. To this end, we find the tangent plane of $\operatorname{CH}(\mathcal{S})$ at $q_0$, which is denoted as $\mathscr{P}$ in the figure. Due to the property of convex hulls, $q^*$, $q_0$, and $p_0$ should lie in the same side of the plane $\mathscr{P}$. Therefore, within $\triangle q^*q_0p_0$, $\angle q^*q_0p_0 \leq 90^\circ$, and we thus have:

\begin{equation}
r^2 + s^2  \geq t^2
\end{equation}

Since $r \leq \operatorname{R_v}(\mathcal{S})$ and $s \leq \operatorname{H_b}(\mathcal{S})$, we have:
\begin{equation}
    t \leq \sqrt{r^2+s^2} \leq \sqrt{\operatorname{R_v}(\mathcal{S})^2 + \operatorname{H_b}(\mathcal{S})^2} \leq \sqrt{2} \max(\operatorname{R_v}(\mathcal{S}), \operatorname{H_b}(\mathcal{S}))
\end{equation}

Moreover, since $p^*$ is $q^*$'s nearest point on the boundary surface of $\mathcal{S}$, we know that $\operatorname{H_i}(\mathcal{S}) = d(p^*, q^*) \leq t$. The theorem thus holds.

\section{Detailed version of Table 1}

\setlength{\tabcolsep}{2pt}
\begin{table}[t]
  \centering
  \footnotesize
  \caption{Quantitative comparison on the PartNet-Mobility dataset~\cite{xiang2020sapien}. ``Better ratio'' indicates the percentage of cases where our method outperforms the baseline.}
    \begin{tabular}{c|cccc|cccc}
    \toprule
          & \multicolumn{4}{c|}{HACD~\cite{mamou2009simple}}     & \multicolumn{4}{c}{V-HACD~\cite{mamou2016volumetric}} \\
    \midrule
          & \multicolumn{2}{c}{\textbf{\# components $\downarrow$}} & \multicolumn{2}{c|}{concavity $\downarrow$} & \multicolumn{2}{c}{\textbf{\# components $\downarrow$}} & \multicolumn{2}{c}{concavity $\downarrow$} \\
          & \multicolumn{1}{c}{theirs} & \multicolumn{1}{c}{ours} & \multicolumn{1}{c}{theirs} & \multicolumn{1}{c|}{ours} & \multicolumn{1}{c}{theirs} & \multicolumn{1}{c}{ours} & \multicolumn{1}{c}{theirs} & \multicolumn{1}{c}{ours} \\
    \midrule
    average   & 33.5  & \textbf{7.3} & 0.414  & \textbf{0.204 } & 44.6 & \textbf{20.1} & 0.055  & \textbf{0.052 } \\
    median & 27    & \textbf{1} & 0.218  & \textbf{0.117 } & 20    & \textbf{13} & 0.045  & \textbf{0.041 } \\
    better ratio &   -   & 90.85\% &-   & 89.18\% &   -   & 96.50\% &   -   & 96.90\% \\
    \bottomrule
    \end{tabular} 
  \label{tab:sapien_full}
\end{table}

\setlength{\tabcolsep}{3pt}
\begin{table*}[t]
  \centering
  \footnotesize
  \caption{ Quantitative comparison on the V-HACD dataset~\cite{mamou2016volumetric}. ``Better ratio'' indicates the percentage of cases where our method outperforms the baseline. }
    \begin{tabular}{c|cccc|cccc|cccc}
    \toprule
          & \multicolumn{4}{c|}{HACD~\cite{mamou2009simple}}     & \multicolumn{4}{c|}{V-HACD~\cite{mamou2016volumetric}}    & \multicolumn{4}{c}{Animation~\cite{thul2018approximate}} \\
    \midrule
          & \multicolumn{2}{c}{\textbf{\# components $\downarrow$}} & \multicolumn{2}{c|}{concavity $\downarrow$} & \multicolumn{2}{c}{\textbf{\# components $\downarrow$}} & \multicolumn{2}{c|}{concavity $\downarrow$} & \multicolumn{2}{c}{\# components $\downarrow$} & \multicolumn{2}{c}{\textbf{concavity $\downarrow$}} \\
          & \multicolumn{1}{c}{theirs} & \multicolumn{1}{c}{ours} & \multicolumn{1}{c}{theirs} & \multicolumn{1}{c|}{ours} & \multicolumn{1}{c}{theirs} & \multicolumn{1}{c}{ours} & \multicolumn{1}{c}{theirs} & \multicolumn{1}{c|}{ours} & \multicolumn{1}{c}{theirs} & \multicolumn{1}{c}{ours} & \multicolumn{1}{c}{theirs} & \multicolumn{1}{c}{ours} \\
    \midrule
    block & 52    & \textbf{2} & \textbf{0.307} & 0.316  & 19    & \textbf{18} & 0.043  & \textbf{0.030} & 18    & 18    & 0.066  & \textbf{0.035} \\
    bunny & 58    & \textbf{53} & 0.050  & \textbf{0.033} & 21    & \textbf{12} & 0.100  & \textbf{0.078} & 52    & 52    & 0.083  & \textbf{0.051} \\
    camel & 64    & \textbf{39} & 0.039  & \textbf{0.023} & 42    & \textbf{21} & 0.077  & \textbf{0.050} & 35    & 35    & 0.061  & \textbf{0.050} \\
    casting & 86    & \textbf{4} & 0.312  & \textbf{0.267} & 59    & \textbf{52} & 0.064  & \textbf{0.037} & 68    & 69    & 0.066  & \textbf{0.056} \\
    chair & 30    & \textbf{6} & 0.183  & \textbf{0.091} & 30    & \textbf{23} & 0.026  & \textbf{0.017} & 13    & 13    & 0.055  & \textbf{0.045} \\
    cow1  & 66    & \textbf{45} & 0.038  & \textbf{0.022} & 33    & \textbf{20} & 0.062  & \textbf{0.045} & 27    & 27    & 0.059  & \textbf{0.049} \\
    cow2  & \textbf{54} & 68    & 0.021  & \textbf{0.015} & 29    & \textbf{25} & 0.054  & \textbf{0.029} & 25    & 25    & \textbf{0.041} & 0.047  \\
    crank & 90    & \textbf{12} & 0.190  & \textbf{0.097} & 114   & \textbf{12} & 0.187  & \textbf{0.097} & 80    & 80    & 0.186  & \textbf{0.050} \\
    cup   & 65    & \textbf{10} & 0.135  & \textbf{0.083} & 38    & \textbf{23} & 0.075  & \textbf{0.058} & 46    & 47    & 0.315  & \textbf{0.054} \\
    dancer2 & \textbf{34} & 51    & 0.015  & \textbf{0.012} & 49    & \textbf{25} & 0.024  & \textbf{0.017} & 7     & 7     & 0.040  & \textbf{0.039} \\
    deer\_bound & 55    & \textbf{44} & 0.040  & \textbf{0.021} & 69    & \textbf{44} & 0.037  & \textbf{0.021} & 35    & 35    & \textbf{0.044} & 0.047  \\
    dilo  & \textbf{42} & 45    & 0.016  & \textbf{0.011} & 35    & \textbf{29} & 0.027  & \textbf{0.015} & 15    & 15    & 0.059  & \textbf{0.051} \\
    dino  & \textbf{51} & 70    & 0.019  & \textbf{0.013} & 32    & \textbf{30} & 0.048  & \textbf{0.024} & 25    & 25    & \textbf{0.039} & 0.053  \\
    DRAGON\_F & 90    & \textbf{29} & 0.085  & \textbf{0.052} & 76    & \textbf{42} & 0.064  & \textbf{0.040} & 52    & 52    & 0.061  & \textbf{0.058} \\
    drum  & 17    & \textbf{12} & 0.059  & \textbf{0.035} & 6     & \textbf{5} & 0.100  & \textbf{0.054} & 16    & 16    & 0.068  & \textbf{0.047} \\
    egea  & \textbf{34} & 52    & 0.032  & \textbf{0.023} & \textbf{6} & \textbf{6} & 0.102  & \textbf{0.074} & 26    & 26    & \textbf{0.048} & 0.050  \\
    eight & 42    & \textbf{37} & 0.019  & \textbf{0.014} & 26    & \textbf{18} & 0.033  & \textbf{0.024} & 17    & 17    & 0.041  & \textbf{0.040} \\
    elephant & 84    & \textbf{52} & 0.043  & \textbf{0.029} & 62    & \textbf{43} & 0.053  & \textbf{0.034} & 45    & 45    & 0.060  & \textbf{0.051} \\
    elk   & \textbf{42} & 46    & 0.052  & \textbf{0.032} & 40    & \textbf{28} & 0.098  & \textbf{0.058} & 52    & 52    & \textbf{0.045} & 0.050  \\
    face-YH & 125   & \textbf{10} & 0.240  & \textbf{0.160} & 240   & \textbf{113} & 0.039  & \textbf{0.027} & 82    & 82    & 0.067  & \textbf{0.050} \\
    feline & 91    & \textbf{27} & 0.086  & \textbf{0.054} & 87    & \textbf{29} & 0.079  & \textbf{0.051} & 54    & 54    & \textbf{0.051} & 0.052  \\
    fish  & 26    & \textbf{8} & 0.081  & \textbf{0.047} & 17    & \textbf{10} & 0.072  & \textbf{0.042} & 13    & 13    & \textbf{0.044} & 0.045  \\
    foot  & \textbf{30} & 32    & 0.018  & \textbf{0.015} & \textbf{6} & \textbf{6} & 0.040  & \textbf{0.034} & 5     & 5     & \textbf{0.041} & 0.050  \\
    genus3 & 53    & \textbf{31} & 0.034  & \textbf{0.023} & 29    & \textbf{16} & 0.064  & \textbf{0.046} & 23    & 23    & 0.054  & \textbf{0.048} \\
    greek\_sculpture & 83    & \textbf{71} & 0.029  & \textbf{0.021} & 47    & \textbf{19} & 0.059  & \textbf{0.047} & 30    & 30    & 0.054  & \textbf{0.050} \\
    Hand1 & 49    & \textbf{41} & 0.035  & \textbf{0.024} & 27    & \textbf{16} & 0.078  & \textbf{0.051} & 26    & 26    & 0.061  & \textbf{0.048} \\
    hand2 & 49    & \textbf{41} & 0.038  & \textbf{0.024} & 27    & \textbf{17} & 0.077  & \textbf{0.048} & 25    & 25    & 0.060  & \textbf{0.047} \\
    helix & 37    & \textbf{36} & 0.016  & \textbf{0.013} & \textbf{32} & \textbf{32} & 0.027  & \textbf{0.016} & 21    & 21    & \textbf{0.040} & 0.047  \\
    helmet & 28    & \textbf{3} & 0.211  & \textbf{0.090} & 7     & \textbf{5} & 0.103  & \textbf{0.082} & 10    & 10    & \textbf{0.068} & 0.087  \\
    hero  & 118   & \textbf{13} & 0.144  & \textbf{0.097} & 228   & \textbf{51} & 0.060  & \textbf{0.044} & 78    & 78    & 0.090  & \textbf{0.052} \\
    homer & \textbf{61} & 66    & 0.020  & \textbf{0.014} & 23    & \textbf{17} & 0.044  & \textbf{0.030} & 16    & 16    & 0.062  & \textbf{0.047} \\
    hornbug & 114   & \textbf{51} & 0.059  & \textbf{0.038} & 120   & \textbf{38} & 0.076  & \textbf{0.048} & 67    & 67    & 0.059  & \textbf{0.050} \\
    horse & 55    & \textbf{45} & 0.033  & \textbf{0.020} & 32    & \textbf{22} & 0.055  & \textbf{0.037} & 24    & 24    & 0.061  & \textbf{0.046} \\
    maneki-neko & 130   & \textbf{1} & 0.394  & \textbf{0.299} & 516   & \textbf{278} & 0.038  & \textbf{0.027} & 190   & 191   & 0.096  & \textbf{0.051} \\
    mannequin-devil & 50    & \textbf{13} & 0.085  & \textbf{0.062} & 8     & \textbf{3} & 0.190  & \textbf{0.116} & 36    & 36    & 0.100  & \textbf{0.051} \\
    mannequin & \textbf{42}    & \textbf{42} & 0.037  & \textbf{0.022} & 9     & \textbf{8} & 0.101  & \textbf{0.062} & 23    & 23    & 0.062  & \textbf{0.052} \\
    mask  & 112   & \textbf{6} & 0.259  & \textbf{0.178} & 183   & \textbf{77} & 0.036  & \textbf{0.026} & 50    & 50    & 0.069  & \textbf{0.060} \\
    moaimoai & \textbf{53} & 63    & 0.022  & \textbf{0.016} & 8     & \textbf{6} & 0.083  & \textbf{0.066} & 17    & 17    & 0.103  & \textbf{0.048} \\
    monk  & 75    & \textbf{55} & 0.036  & \textbf{0.024} & 23    & \textbf{7} & 0.108  & \textbf{0.078} & 29    & 29    & 0.095  & \textbf{0.051} \\
    octopus & 81    & \textbf{48} & 0.058  & \textbf{0.032} & 96    & \textbf{65} & 0.041  & \textbf{0.023} & 54    & 54    & 0.065  & \textbf{0.053} \\
    pig   & \textbf{46} & 60    & 0.022  & \textbf{0.016} & 13    & \textbf{11} & 0.074  & \textbf{0.053} & 18    & 18    & \textbf{0.046} & 0.050  \\
    pinocchio\_b & 150   & \textbf{4} & 0.319  & \textbf{0.249} & 330   & \textbf{140} & 0.048  & \textbf{0.032} & 132   & 132   & 0.069  & \textbf{0.050} \\
    polygirl & 55    & \textbf{15} & 0.063  & \textbf{0.043} & 30    & \textbf{17} & 0.059  & \textbf{0.039} & 23    & 23    & 0.071  & \textbf{0.060} \\
    rabbit & 36    & \textbf{5} & 0.118  & \textbf{0.075} & 11    & \textbf{7} & 0.076  & \textbf{0.055} & 15    & 15    & \textbf{0.048} & 0.050  \\
    rocker-arm & 65    & \textbf{26} & 0.062  & \textbf{0.033} & 51    & \textbf{23} & 0.069  & \textbf{0.037} & 29    & 29    & 0.074  & \textbf{0.050} \\
    screwdriver & 48    & \textbf{38} & 0.025  & \textbf{0.016} & 27    & \textbf{26} & 0.035  & \textbf{0.022} & 18    & 18    & \textbf{0.038} & 0.046  \\
    shark\_b & 84    & \textbf{7} & 0.098  & \textbf{0.056} & 336   & \textbf{80} & 0.013  & \textbf{0.010} & 16    & 16    & \textbf{0.040} & 0.052  \\
    Sketched-Brunnen & 101   & \textbf{20} & 0.145  & \textbf{0.090} & 110   & \textbf{67} & 0.047  & \textbf{0.033} & 65    & 65    & 0.062  & \textbf{0.050} \\
    sledge & 30    & \textbf{2} & 0.277  & \textbf{0.238} & 24    & \textbf{18} & 0.060  & \textbf{0.021} & 18    & 18    & 0.034  & \textbf{0.031} \\
    squirrel & \textbf{44} & \textbf{44} & 0.048  & \textbf{0.034} & 14    & \textbf{10} & 0.113  & \textbf{0.079} & 46    & 46    & 0.058  & \textbf{0.053} \\
    sword & \textbf{15} & 67    & 0.025  & \textbf{0.016} & 32    & \textbf{9} & 0.055  & \textbf{0.040} & 14    & 14    & \textbf{0.030} & 0.049  \\
    table & \textbf{5} & 15    & 0.012  & \textbf{0.007} & \textbf{6} & 9     & 0.039  & \textbf{0.014} & 7     & 7     & 0.329 & \textbf{0.048}  \\
    Teapot & 83    & \textbf{7} & 0.240  & \textbf{0.216} & 61    & \textbf{23} & 0.152  & \textbf{0.091} & 63    & 63    & \textbf{0.085} & 0.097  \\
    test2 & 27    & \textbf{1} & 0.651  & \textbf{0.521} & 15    & \textbf{13} & 0.058  & \textbf{0.043} & 21    & 21    & 0.075  & \textbf{0.046} \\
    test  & 8     & \textbf{1} & \textbf{0.534} & \textbf{0.534} & \textbf{2} & \textbf{2} & 0.058  & \textbf{0.029} & 3     & 3     & \textbf{0.011} & 0.013  \\
    torus & 34    & \textbf{4} & 0.227  & \textbf{0.132} & 11    & \textbf{9} & 0.055  & \textbf{0.038} & 12    & 12    & \textbf{0.037} & 0.048  \\
    tstTorusModel3 & 36    & \textbf{4} & 0.250  & \textbf{0.179} & \textbf{16} & \textbf{16} & 0.037  & \textbf{0.024} & 12    & 12    & \textbf{0.042} & \textbf{0.042} \\
    tstTorusModel & 34    & \textbf{4} & 0.226  & \textbf{0.132} & 11    & \textbf{9} & 0.055  & \textbf{0.042} & 11    & 11    & \textbf{0.042} & 0.047  \\
    tube1 & 8     & \textbf{4} & 0.223  & \textbf{0.031} & \textbf{4} & \textbf{4} & 0.050  & \textbf{0.031} & 4     & 4     & 0.042  & \textbf{0.032} \\
    venus-original & \textbf{42} & 44    & 0.036  & \textbf{0.026} & 7     & \textbf{6} & 0.100  & \textbf{0.072} & 27    & 27    & 0.056  & \textbf{0.050} \\
    venus & \textbf{46} & 52    & 0.024  & \textbf{0.017} & 13    & \textbf{10} & 0.075  & \textbf{0.050} & 20    & 20    & 0.056  & \textbf{0.048} \\
    \midrule
    average   & 57.6  & \textbf{29.6} & 0.118  & \textbf{0.084} & 60.2  & \textbf{29.8} & 0.067  & \textbf{0.044} & 34.4  & 34.5  & 0.069  & \textbf{0.049} \\
    median & 51    & \textbf{31} & 0.058  & \textbf{0.033} & 29    & \textbf{18} & 0.059  & \textbf{0.040} & 25    & 25    & 0.059  & \textbf{0.050} \\
    better ratio &  -   & 77.05\% &  -   & 98.36\% &  -    & 98.36\% &   -   & 100.00\% &   -   & 95.10\% &  -    & 81.97\% \\
    \bottomrule
    \end{tabular} 
  \label{tab:vhacd_full}
\end{table*}

We show the complete quantitative comparison (for all objects) of the V-HACD dataset~\cite{mamou2016volumetric} in Table~\ref{tab:vhacd_full}. Since the PartNet-Mobility dataset~\cite{xiang2020sapien} contains thousands of objects, we only report insightful statistics in Table~\ref{tab:sapien_full}. We compare each baseline algorithm with our method separately. For both HACD~\cite{mamou2009simple} and V-HACD~\cite{mamou2016volumetric}, we let our method produce decomposition results with lower concavity scores and aim to compare the numbers of decomposed components. For Animation~\cite{thul2018approximate}, we aim to match the numbers of decomposed components and compare the concavity scores. In addition to the average and median, we also calculate the percentage of cases where our method outperforms the baseline (denoted as ``better ratio'').

\section{Default value of hyper-parameters}

The default value of the hyper-parameters are $m = 20$, $t = 500$, $d = 4$, $k = 0.3$. We sample 3,000 points per unit area when computing $\operatorname{H_b}$. The hyper-parameters are consistent across datasets and experiments except for ablating the hyper-parameter. In experiments, all of our methods include the merging stage unless otherwise noted.

\section{Ablation study of the exploration parameter $c$}

\begin{figure}[t]
\begin{center}
  \includegraphics[width=0.88\linewidth]{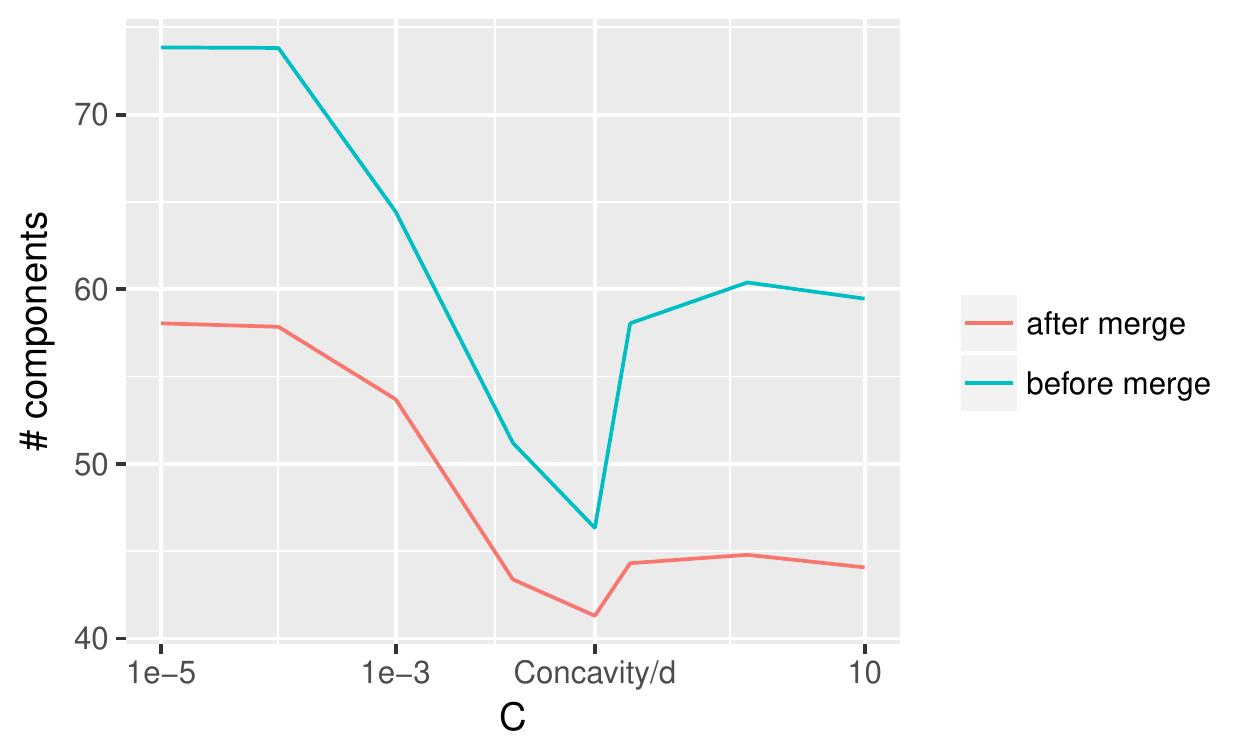}
\end{center}
  \caption{The influence of the exploration parameter $c$.}
\label{fig:c_ablation}
\end{figure}

In the tree search, we use the UCB (Upper Confidence Bound)~\cite{kocsis2006bandit} term to select a tree node for expansion:

\begin{equation}
    \operatorname{Q}(n) + c \sqrt{\frac{2\ln{\operatorname{N}(n^{\prime})}}{\operatorname{N}(n)}}
\end{equation}

where $c$ is the parameter balancing the exploration and exploitation. We study the influence of $c$ on the V-HACD dataset and report the results in Figure~\ref{fig:c_ablation}. As shown in the figure, when $c$ is set to 0, the UCB term only uses the existing value function $Q(n)$ to select a node, and no exploration occurs. In this case, MCTS almost degenerates into a one-step greedy search, and the resulting number of components increases a lot. In contrast, when $c$ is set to a large number, the UCB term ignores the influence of the quality function, and the MCTS degrades to an inefficient exhaustive search algorithm, which also leads to sub-optimal results. Instead, we empirically set $c$ to be $\widetilde{\operatorname{Concavity}}(\mathcal{S})/d$, where $\widetilde{\operatorname{Concavity}}(\mathcal{S})$ is the concavity score of each individual input component, and $d$ is the depth of the tree search. We find that it generates good results in general.

\section{How well does $\operatorname{R_v}$ approximates $\operatorname{H_i}$?}

In addition to the theoretical proof, we experimentally verify that Theorem 1 holds for all shapes on both datasets. We empirically set $k$ to 0.3 by calculating the average $\frac{\operatorname{H_i}}{\operatorname{R_v}}$. We find that $\max(\operatorname{H_b}, k\operatorname{R_v})$ approximates $\max(\operatorname{H_b}, \operatorname{H_i})$  well in practice, with a small absolute or relative error. Specifically, we compare two metrics on the PartNetM dataset (13,536 shapes), and find that for more than 94\% shapes, we have: $\left|\max \left(\operatorname{H_{b}}, k \operatorname{R_{v}}\right)-\max \left(\operatorname{H_{b}}, \operatorname{H_{i}}\right)\right| \leq 0.02$ or $0.8 \leq \max \left(\operatorname{H_{b}}, k \operatorname{R_{v}}\right) / \max \left(\operatorname{H_{b}}, \operatorname{H_{i}}\right) \leq 1.2$.

\end{document}